\documentstyle[12pt]{article}
\gdef\Feynmanlength{\setlength{\unitlength}{0.01pt}}  

\global\newcount\LINETYPE
\global\newcount\LINEDIRECTION
\global\newcount\LINECONFIGURATION
\newcommand{\LTYPE}{\LINETYPE}
\newcommand{\LDIR}{\LINEDIRECTION}

\global\LINETYPE=1  \global\LINEDIRECTION=0  \global\LINECONFIGURATION=0
\global\newcount\fermion    \fermion=1
\global\newcount\scalar     \scalar=2
\global\newcount\photon     \photon=3
\global\newcount\gluon      \gluon=4
\global\newcount\SPECIAL    \SPECIAL=5
    \gdef\E{2}   
       
\global\newcount\REG            \global\REG=0
\global\newcount\FLIPPED        \global\FLIPPED=1
\global\newcount\CURLY          \global\CURLY=2
\global\newcount\FLIPPEDCURLY   \global\FLIPPEDCURLY=3
\global\newcount\FLAT           \global\FLAT=4
\global\newcount\FLIPPEDFLAT    \global\FLIPPEDFLAT=5
\global\newcount\CENTRAL        \global\CENTRAL=6
\global\newcount\FLIPPEDCENTRAL \global\FLIPPEDCENTRAL=7
             
\global\newcount\SQUASHEDGLUON  \global\SQUASHEDGLUON=8

\newcount\adjx \adjx=0
\newcount\adjy \adjy=0
\global\newdimen\BIGPHOTONS     \BIGPHOTONS=0pt  
\gdef\bigphotons{\global\BIGPHOTONS=12pt}
\global\newdimen\THICKPHOTONS     \THICKPHOTONS=0pt  
\global\newdimen\THICKPHOTONSWITCH    \THICKPHOTONSWITCH=0pt
\gdef\THICKPHOTONTEST{
\THICKPHOTONSWITCH=0pt
\ifdim\THICKPHOTONS=0pt \relax
  \else \ifnum\LTYPE=3
           \ifnum\LDIR=2 \THICKPHOTONSWITCH=1pt \fi 
           \ifnum\LDIR=6 \THICKPHOTONSWITCH=1pt \fi 
        \fi
\fi
}  

\global\newcount\phantomswitch   \global\phantomswitch=0
\global\newcount\stemlength   \global\stemlength=275   
\global\newcount\absstemlength        
\global\newcount\stemlengthx          
\global\newcount\stemlengthy          
\newdimen\FRONTSTEM  \FRONTSTEM=0pt   
\newdimen\BACKSTEM   \BACKSTEM=0pt    
\newdimen\EITHERSTEM \EITHERSTEM=0pt  
\global\newcount\arrowlength                
\global\newdimen\ATTIP   \global\ATTIP=0pt  
\global\newdimen\ATBASE  \global\ATBASE=1pt 
\global\newcount\unitboxnumber  
\global\newcount\unitboxnumberpo  
\global\newcount\particlelengthx  
\gdef\plengthx{\particlelengthx}
\global\newcount\particlelengthy  
\gdef\plengthy{\particlelengthy}
\global\newcount\boxlengthx  
\global\newcount\boxlengthy  
\global\newcount\particleadjustx  
\global\newcount\particleadjusty  
\global\newcount\particlelength   
\global\newcount\particlefrontx
\gdef\pfrontx{\particlefrontx}
\global\newcount\PFRONTx
\global\newcount\particlefronty
\gdef\pfronty{\particlefronty}
\global\newcount\PFRONTy
\global\newcount\particlebackx
\gdef\pbackx{\particlebackx}
\global\newcount\particlebacky
\gdef\pbacky{\particlebacky}
\global\newcount\particlemidx
\gdef\pmidx{\particlemidx}
\global\newcount\particlemidy
\gdef\pmidy{\particlemidy}
\global\newcount\seglength  \global\newcount\gaplength
\global\gaplength=850  
\global\seglength=1416  
\global\newcount\Xone    \global\newcount\Yone    
\global\newcount\Xtwo    \global\newcount\Ytwo    
\global\newcount\Xthree  \global\newcount\Ythree  
\global\newcount\Xfour   \global\newcount\Yfour   
\global\newcount\Xfive   \global\newcount\Yfive   
\global\newcount\Xsix    \global\newcount\Ysix    
\global\newcount\Xseven  \global\newcount\Yseven  
\global\newcount\Xeight  \global\newcount\Yeight  
\newsavebox{\lastline}  
\global\newcount\numlineparts   
\global\newcount\upperlineadjx  \upperlineadjx=0  
\global\newcount\upperlineadjy  \upperlineadjy=0  
\global\newcount\lowerlineadjx  \lowerlineadjx=0  
\global\newcount\lowerlineadjy  \lowerlineadjy=0  
\global\newcount\thirdlineadjx  \thirdlineadjx=0  
\global\newcount\thirdlineadjy  \thirdlineadjy=0  
\global\newcount\fourthlineadjx \fourthlineadjx=0  
\global\newcount\fourthlineadjy \fourthlineadjy=0  
\global\newcount\unitboxwidth   \unitboxwidth=1000
\global\newcount\unitboxheight  \unitboxheight=0  
\global\newcount\numupperunits  \numupperunits=8  
\global\newcount\numlowerunits  \numlowerunits=8  
\global\newcount\numthirdunits  \numthirdunits=8  
\global\newcount\numfourthunits \numfourthunits=8  
\global\newcount\fermioncount   \global\fermioncount=0
\global\newcount\scalarcount    \global\scalarcount=0
\global\newcount\photoncount    \global\photoncount=0
\global\newcount\gluoncount     \global\gluoncount=0
\global\newcount\SPECIALcount   \global\SPECIALcount=0
\global\newcount\vertexcount    \global\vertexcount=-1
\global\newcount\XDIR
\global\newcount\YDIR
\gdef\SETDIR{  
\ifcase\LDIR
     \global\XDIR=0  \global\YDIR=1   
\or  \global\XDIR=1  \global\YDIR=1   
\or  \global\XDIR=1  \global\YDIR=0   
\or  \global\XDIR=1  \global\YDIR=-1  
\or  \global\XDIR=0  \global\YDIR=-1  
\or  \global\XDIR=-1 \global\YDIR=-1  
\or  \global\XDIR=-1 \global\YDIR=0   
\or  \global\XDIR=-1 \global\YDIR=1   
\else\DIRECTERROR
\fi}  
\gdef\moduloeight#1{
\ifnum#1>7 \global\advance #1 by -8
\relax
\moduloeight#1
\relax
\else \relax
\fi}
\gdef\multroothalf#1{\global\multiply #1 by 7071 \global\divide #1 by 10000}
\gdef\negate#1{\global\multiply #1 by -1}

\gdef\slanttest(#1,#2){
\ifodd\LDIR
\multiply #1 by 7071  \divide #1 by 10000
\multiply #2 by 7071  \divide #2 by 10000
\fi
}
\gdef\gslanttest(#1,#2){
\ifodd\LDIR
\multroothalf#1
\multroothalf#2
\fi
}
\gdef\setplength{ 
\global\particlelengthx=\unitboxwidth
\global\particlelengthy=\unitboxheight
\global\multiply \particlelengthx by \unitboxnumber
\global\multiply \particlelengthy by \unitboxnumber
\global\advance \particlelengthx by \particleadjustx
\global\advance \particlelengthy by \particleadjusty
}
\gdef\boxlengthdefault{  
\global\boxlengthx=\plengthx
\global\boxlengthy=\plengthy
\ifnum\plengthx<0 \global\multiply\boxlengthx by -1 \fi
\ifnum\plengthy<0 \global\multiply\boxlengthy by -1 \fi
}
\gdef\rearcoords{  
\global\particlebacky=\particlefronty
\global\particlebackx=\particlefrontx
\global\advance \particlebackx by \particlelengthx
\global\advance \particlebacky by \particlelengthy
}
\gdef\midcoords{  
\global\particlemidy=\particlefronty
\global\particlemidx=\particlefrontx
\global\stemlengthx=\particlelengthx  
\global\stemlengthy=\particlelengthy
\global\divide\stemlengthx by 2
\global\divide\stemlengthy by 2
\global\advance \particlemidx by \stemlengthx
\global\advance \particlemidy by \stemlengthy
}
\gdef\setcoords(#1,#2,#3)(#4,#5,#6)[#7,#8]{
\global\upperlineadjx=#1
\global\lowerlineadjx=#2
\global\thirdlineadjx=#3
\global\upperlineadjy=#4
\global\lowerlineadjy=#5
\global\thirdlineadjy=#6
\global\unitboxwidth=#7
\global\unitboxheight=#8
}
\gdef\drawoldpic#1(#2,#3){  
\global\particlefrontx=#2
\global\particlefronty=#3
\rearcoords
\midcoords
\put(#2,#3){\usebox{#1}}
}
\gdef\drawsavedline`#1' as #2[#3#4](#5,#6)[#7]{
\global\LINETYPE=#2
\global\LINEDIRECTION=#3
\global\LINECONFIGURATION=#4
\global\particlefrontx=#5
\global\particlefronty=#6
\global\unitboxnumber=#7
\selectcase
\rearcoords
\midcoords
\ifnum\phantomswitch=0 \drawas{#1}\fi
}

\gdef\drawas#1{
\global\savebox{#1}(\boxlengthx,\boxlengthy){
\setlength{\unitlength}{0.01pt}
\begin{picture}(\boxlengthx,\boxlengthy)
\multiput(\upperlineadjx,\upperlineadjy)(\unitboxwidth,\unitboxheight)
{\numupperunits}{\upperunitbox}
\ifnum\numlineparts > 1  
\multiput(\lowerlineadjx,\lowerlineadjy)(\unitboxwidth,\unitboxheight)
{\numlowerunits}{\lowerunitbox}
\fi
\ifnum\numlineparts > 2  
\multiput(\thirdlineadjx,\thirdlineadjy)(\unitboxwidth,\unitboxheight)
{\numthirdunits}{\thirdunitbox}
\fi
\ifnum\numlineparts > 3  
\multiput(\fourthlineadjx,\fourthlineadjy)(\unitboxwidth,\unitboxheight)
{\numfourthunits}{\lowerunitbox}
\fi
\end{picture} }
\global\PFRONTx=\pfrontx  \global\PFRONTy=\pfronty   
\SETFRONTSTEM
\THICKPHOTONTEST
\ifdim\THICKPHOTONSWITCH=1pt\global\advance\PFRONTy by 20  \fi
\put(\PFRONTx,\PFRONTy) {\usebox{#1}}   
\ifdim\THICKPHOTONSWITCH=1pt
\global\advance\PFRONTy by -40
\put(\PFRONTx,\PFRONTy) {\usebox{#1}}   
\global\advance \PFRONTy by 20  
\fi  
\SETBACKSTEM
\seglength=1416   \gaplength=850   
}
\gdef\drawandsaveline`#1' as #2[#3#4](#5,#6)[#7]{
\global\newsavebox{#1}
\drawsavedline`#1' as #2[#3#4](#5,#6)[#7]
}

\gdef\drawline#1[#2#3](#4,#5)[#6]{   
\drawsavedline`\lastline' as #1[#2#3](#4,#5)[#6]}

\gdef\TYPEERROR{\message{*** ERROR IN PARTICLE TYPE SELECTION ***}
\message{+++ Try with line type \fermion,\scalar,\photon,\gluon
(see manual) +++}\SETERR}
\gdef\DIRECTERROR{\SETERR\message{*** ERROR IN PARTICLE DIRECTION SELECTION
***}
\message{+++ Try again with direction N, NE, E, SE  etc. or see manual +++}}
\gdef\UNIMPERROR{\message{*** ERROR IN PARTICLE OPTIONS SELECTION ***}
\message{
+++ The requested options combination has not yet been implemented +++}\SETERR}
\gdef\SETERR{\gdef\upperunitbox{{\tiny Error}}  
\gdef\lowerunitbox{\relax}
\gdef\thirdunitbox{\relax}
}
\gdef\neglengthcheck{\ifnum\unitboxnumber < 1
\message{   *** ERROR:  PARTICLE OF NEGATIVE OR ZERO LENGTH REQUESTED. ***   }
\message{   ***         TAKING ABSOLUTE VALUE. ***   }\negate\unitboxnumber
\fi}
\gdef\selectcase{
\neglengthcheck   
\SETDIR
\ifcase\LINETYPE
\TYPEERROR  
\or \selectfermion  
\or \selectscalar   
\or \selectphoton   
\or \selectgluon    
\or \selectspecial  
\else \TYPEERROR \fi  }
\gdef\selectfermion{
\ifnum\fermioncount=0 \input FERMIONSETUP \fi
\global\advance\fermioncount by 1  
\ALLfermion
}
\gdef\selectscalar{
\ifnum\scalarcount=0 \input SCALARSETUP \fi
\global\advance\scalarcount by 1  
\ALLscalar
}
\gdef\selectphoton{   
\ifnum\photoncount=0 
\newcount\numwiggles    \newcount\numwigglespo
\global\newcount\photonlengthx
\global\newcount\photonlengthy
\global\newcount\photonfrontx  
\global\newcount\photonfronty  
\global\newcount\photonbackx
\global\newcount\photonbacky
\newcount\halfwigglelength
\global\font\Twelverom=cmr12
\global\font\Tenrom=cmr10
\gdef\Lbr{{\Twelverom(}}   \gdef\Rbr{{\Twelverom)}}
\gdef\SLbr{{\Tenrom(}}     \gdef\SRbr{{\Tenrom)}}
\gdef\Smile{{\large$\smile$}}  
\gdef\Frown{{\large$\frown$}}  
\ifdim\BIGPHOTONS>0pt  \gdef\Smile{$\smile$} \gdef\Frown{$\frown$} \fi
\gdef\selectphoton{   
\global\advance\photoncount by 1  
\global\photonfrontx=\particlefrontx   
\global\photonfronty=\particlefronty   
\ifnum\unitboxnumber > 50
\message{   *** WARNING *** Photon with
\the\unitboxnumber\space half-wiggles requested ***   }
\ifnum\unitboxnumber > 150
\message{   *** Reducing photon length to 10 half-wiggles (max 150) ***   }
\ifnum\unitboxnumber > 1000
\message{   *** Probable Cause:  Photon selected instead of Fermion ***   }
\fi \global\unitboxnumber=10 \fi \fi  
\numwiggles=\unitboxnumber
\divide\numwiggles by 2
\global\unitboxnumberpo=\numwiggles 
\global\multiply \unitboxnumberpo by -1
\numwigglespo=\unitboxnumber
\advance\numwigglespo by \unitboxnumberpo 
\global\numlineparts = 2  
\global\numupperunits=\numwigglespo  
\global\numlowerunits=\numwiggles  
\particleadjustx=0  
\particleadjusty=0  
\ifcase\LINEDIRECTION
     \Nphoton    
\or  \NEphoton   
\or  \Ephoton    
\or  \SEphoton   
\or  \Sphoton    
\or  \SWphoton   
\or  \Wphoton    
\or  \NWphoton   
\else\DIRECTERROR \fi
\setplength
\global\divide\plengthx by 2  \global\divide\plengthy by 2
\rearcoords  \boxlengthdefault   \midcoords
\global\photonbackx=\pbackx  
\global\photonbacky=\pbacky  
\global\photonlengthx=\plengthx  
\global\photonlengthy=\plengthy  
}
\gdef\SETUNITBOX(#1)[#2][#3]{ 
\gdef\upperunitbox{\oval(#1,#1)[#2]}
\gdef\lowerunitbox{\oval(#1,#1)[#3]}
}
\gdef\Nphoton{  
\ifcase\LINECONFIGURATION  
\setcoords(-490,-250,0)(260,1250,0)[0,2000]
\gdef\upperunitbox{\SLbr}   \gdef\lowerunitbox{\SRbr}
\particleadjusty=10
\or 
\setcoords(-271,-501,0)(250,1250,0)[0,2000]
\gdef\upperunitbox{\SRbr}   \gdef\lowerunitbox{\SLbr}
\or 
\particleadjusty=0
\setcoords(-501,-351,0)(300,1400,0)[0,2200]
\gdef\upperunitbox{\Lbr}   \gdef\lowerunitbox{\Rbr}
\or 
\setcoords(-353,-499,0)(300,1400,0)[0,2200]
\gdef\upperunitbox{\Rbr}   \gdef\lowerunitbox{\Lbr}
\or 
\setcoords(-481,-371,0)(280,1300,0)[0,2000]
\gdef\upperunitbox{\Lbr}   \gdef\lowerunitbox{\Rbr}
\particleadjusty=150
\ifnum\numwiggles=\number\numwigglespo \particleadjustx=-50 \fi
\or 
\setcoords(-321,-391,0)(280,1300,0)[0,2000]
\gdef\upperunitbox{\Rbr}   \gdef\lowerunitbox{\Lbr}
\particleadjusty=150
\ifnum\numwiggles=\number\numwigglespo \particleadjustx=80 \fi
\or 
\setcoords(-490,-260,0)(300,1500,0)[0,2400]
\gdef\upperunitbox{\Lbr}   \gdef\lowerunitbox{\Rbr}
\or 
\setcoords(-301,-531,0)(300,1500,0)[0,2400]
\gdef\upperunitbox{\Rbr}   \gdef\lowerunitbox{\Lbr}
\else \UNIMPERROR
\fi
}
\gdef\NEphoton{    
\ifcase\LINECONFIGURATION  
\setcoords(425,425,0)(1250,0,0)[1250,1250]       \SETUNITBOX(1250)[br][tl]
\ifnum\numwigglespo > \number \numwiggles \particleadjustx=15 \fi
\or 
\setcoords(1050,-200,0)(625,625,0)[1250,1250]    \SETUNITBOX(1250)[tl][br]
\ifnum\numwigglespo > \number \numwiggles \particleadjustx=25 \fi
\or 
\setcoords(500,500,0)(1400,0,0)[1400,1400]       \SETUNITBOX(1400)[br][tl]
\or 
\setcoords(1200,-200,0)(700,700,0)[1400,1400]    \SETUNITBOX(1400)[tl][br]
\or 
\setcoords(400,400,0)(1200,0,0)[1200,1200]       \SETUNITBOX(1200)[br][tl]
\or 
\setcoords(1000,-200,0)(600,600,0)[1200,1200]    \SETUNITBOX(1200)[tl][br]
\else \UNIMPERROR
\fi
\numupperunits=\numwiggles   \numlowerunits=\numwigglespo
}
\gdef\Ephoton{    
\ifcase\LINECONFIGURATION  
\setcoords(-285,715,0)(-150,-400,0)[2005,0]
\gdef\upperunitbox{\Frown}   \gdef\lowerunitbox{\Smile}
\or  
\setcoords(-285,715,0)(-420,-170,0)[2005,0]
\gdef\upperunitbox{\Smile}   \gdef\lowerunitbox{\Frown}
\else \UNIMPERROR
\fi
\particleadjustx=-15 
}
\gdef\SEphoton{   
\ifcase\LINECONFIGURATION  
\setcoords(-200,1050,0)(-625,-625,0)[1250,-1250] \SETUNITBOX(1250)[tr][bl]
\ifnum\numwigglespo > \number \numwiggles \particleadjustx=25 \fi
\or 
\setcoords(425,425,0)(0,-1250,0)[1250,-1250]     \SETUNITBOX(1250)[bl][tr]
\ifnum\numwigglespo > \number \numwiggles \particleadjustx=15 \fi
\or 
\setcoords(-200,1200,0)(-700,-700,0)[1400,-1400] \SETUNITBOX(1400)[tr][bl]
\or 
\setcoords(500,500,0)(0,-1400,0)[1400,-1400]     \SETUNITBOX(1400)[bl][tr]
\or 
\setcoords(-200,1000,0)(-600,-600,0)[1200,-1200] \SETUNITBOX(1200)[tr][bl]
\particleadjustx=-20
\or 
\setcoords(420,420,0)(0,-1200,0)[1200,-1200]     \SETUNITBOX(1200)[bl][tr]
\particleadjustx=40
\else \UNIMPERROR
\fi
}
\gdef\Sphoton{  
\ifcase\LINECONFIGURATION  
\setcoords(-252,-490,0)(-740,-1740,0)[0,-2000]
\gdef\upperunitbox{\SRbr}   \gdef\lowerunitbox{\SLbr}
\or 
\setcoords(-490,-260,0)(-740,-1740,0)[0,-2002]
\gdef\upperunitbox{\SLbr}   \gdef\lowerunitbox{\SRbr}
\or 
\setcoords(-299,-449,0)(-870,-1970,0)[0,-2200]
\gdef\upperunitbox{\Rbr}    \gdef\lowerunitbox{\Lbr}
\particleadjusty=-95
\or 
\setcoords(-517,-371,0)(-900,-2000,0)[0,-2200]
\gdef\upperunitbox{\Lbr}    \gdef\lowerunitbox{\Rbr}
\particleadjusty=-165
\or 
\setcoords(-299,-409,0)(-885,-1905,0)[0,-2000]
\gdef\upperunitbox{\Rbr}   \gdef\lowerunitbox{\Lbr}
\particleadjustx=50     \particleadjusty=-380
\ifodd\unitboxnumber\relax\else\particleadjustx=250 \particleadjusty=-400 \fi
\or 
\setcoords(-519,-449,0)(-900,-1920,0)[0,-2000]
\gdef\upperunitbox{\Lbr}   \gdef\lowerunitbox{\Rbr}
\particleadjusty=-370
\ifodd\unitboxnumber\relax\else\particleadjustx=-240 \particleadjusty=-400 \fi
\or 
\gdef\upperunitbox{\Rbr}   \gdef\lowerunitbox{\Lbr}
\setcoords(-325,-555,0)(-900,-2100,0)[0,-2400]
\particleadjusty=-40
\or 
\setcoords(-505,-275,0)(-900,-2100,0)[0,-2400]
\gdef\upperunitbox{\Lbr}   \gdef\lowerunitbox{\Rbr}
\particleadjusty=-30  
\else \UNIMPERROR
\fi
}
\gdef\SWphoton{  
\ifcase\LINECONFIGURATION  
\setcoords(-825,-825,0)(0,-1250,0)[-1250,-1250]     \SETUNITBOX(1250)[br][tl]
\or 
\setcoords(-175,-1425,0)(-625,-625,0)[-1250,-1250]  \SETUNITBOX(1250)[tl][br]
\or 
\setcoords(-900,-900,0)(0,-1410,0)[-1400,-1400]     \SETUNITBOX(1400)[br][tl]
\or 
\setcoords(-200,-1600,0)(-700,-700,0)[-1400,-1400]  \SETUNITBOX(1400)[tl][br]
\or 
\setcoords(-800,-800,0)(0,-1200,0)[-1200,-1200]     \SETUNITBOX(1200)[br][tl]
\or 
\setcoords(-200,-1400,0)(-600,-600,0)[-1200,-1200]  \SETUNITBOX(1200)[tl][br]
\else \UNIMPERROR
\fi
}
\gdef\Wphoton{
\ifcase\LINECONFIGURATION 
\setcoords(-2245,-1245,0)(-150,-400,0)[-2005,0]
\gdef\upperunitbox{\Frown}   \gdef\lowerunitbox{\Smile}
\or 
\setcoords(-2245,-1245,0)(-400,-150,0)[-2005,0]
\gdef\upperunitbox{\Smile}   \gdef\lowerunitbox{\Frown}
\else \UNIMPERROR
\fi
\particleadjustx=57 
\ifnum\numwigglespo=\number\numwiggles \particleadjustx=0  \fi
\numlowerunits=\numwigglespo   \numupperunits=\numwiggles
}
\gdef\NWphoton{  
\ifcase\LINECONFIGURATION  
\setcoords(-200,-1425,0)(625,625,0)[-1250,1250]   \SETUNITBOX(1250)[bl][tr]
\or 
\setcoords(-825,-825,0)(0,1250,0)[-1250,1250]     \SETUNITBOX(1250)[tr][bl]
\ifnum\numwigglespo > \number \numwiggles \particleadjusty=-15 \fi
\or 
\setcoords(-200,-1600,0)(700,700,0)[-1400,1400]   \SETUNITBOX(1400)[bl][tr]
\or 
\setcoords(-900,-900,0)(0,1400,0)[-1400,1400]     \SETUNITBOX(1400)[tr][bl]
\or 
\setcoords(-200,-1400,0)(600,600,0)[-1200,1200]   \SETUNITBOX(1200)[bl][tr]
\or 
\setcoords(-800,-800,0)(0,1200,0)[-1200,1200]     \SETUNITBOX(1200)[tr][bl]
\else \UNIMPERROR
\fi
}
  \fi
\selectphoton
}
\gdef\selectgluon{   
\ifnum\gluoncount=0 \input GLUONSETUP  \fi
\selectgluon
}
\gdef\selectspecial{\UNIMPERROR}
\gdef\checkvertex{ 
\ifnum\vertexcount=-1   \input VERTEX  \fi}
\gdef\drawvertex#1[#2#3](#4,#5)[#6]{\checkvertex\drawvertex#1[#2#3](#4,#5)[#6]}
\gdef\vertexcap#1{\checkvertex\vertexcap#1}
\gdef\vertexcaps{\checkvertex\vertexcaps}
\gdef\vertexlink#1{\checkvertex\vertexlink#1}
\gdef\vertexlinks{\checkvertex\vertexlinks}
\gdef\stemvertex#1{\checkvertex\stemvertex#1}
\gdef\stemvertices{\checkvertex\stemvertices}
\gdef\flipvertex{\checkvertex\flipvertex}
\global\arrowlength=349  
\gdef\drawarrow[#1#2](#3,#4){
\global\LDIR=#1
\SETDIR
\global\boxlengthx=#3  
\global\boxlengthy=#4  
\ifdim#2=1pt  
\adjx=\arrowlength      \adjy=\arrowlength
\multiply\adjx by \XDIR \multiply\adjy by \YDIR  
\slanttest(\adjx,\adjy)
\global\advance\boxlengthx by \adjx    \global\advance\boxlengthy by \adjy
\fi
\ifnum\phantomswitch=0\put(\boxlengthx,\boxlengthy){\vector(\XDIR,\YDIR){0}}\fi
}  
\gdef\SETFRONTSTEM{
\EITHERSTEM=\FRONTSTEM   \advance\EITHERSTEM by \BACKSTEM
\ifdim\EITHERSTEM>0pt
\global\stemlengthx=\stemlength   \global\stemlengthy=\stemlength
\global\absstemlength=\stemlength
\SETDIR
\gslanttest(\stemlengthx,\stemlengthy)
\gslanttest(\absstemlength,\REG)  
\ifnum\XDIR=0 \stemlengthx=0 \fi
\ifnum\YDIR=0 \stemlengthy=0 \fi
\global\multiply\stemlengthx by \XDIR
\global\multiply\stemlengthy by \YDIR
\ifdim\FRONTSTEM=1pt
\ifnum\phantomswitch=0
          \put(\pfrontx,\pfronty){\line(\XDIR,\YDIR){\absstemlength}}\fi
\global\advance\plengthx by \stemlengthx
\global\advance\plengthy by \stemlengthy
\global\advance\PFRONTx by \stemlengthx
\global\advance\PFRONTy by \stemlengthy
\global\advance\pmidx by \stemlengthx
\global\advance\pmidy by \stemlengthy
\global\advance\pbackx by \stemlengthx
\global\advance\pbacky by \stemlengthy
\ifnum\LTYPE=3
\global\photonfrontx=\PFRONTx  \global\photonfronty=\PFRONTy
\global\photonbackx=\pbackx    \global\photonbacky=\pbacky
\fi  
\ifnum\LTYPE=4
\global\gluonfrontx=\PFRONTx  \global\gluonfronty=\PFRONTy
\global\gluonbackx=\pbackx    \global\gluonbacky=\pbacky
\fi  
\fi  
\fi  
}    
\gdef\SETBACKSTEM{
\ifdim\BACKSTEM=1pt
\ifnum\phantomswitch=0
       \put(\pbackx,\pbacky){\line(\XDIR,\YDIR){\absstemlength}}\fi
\global\advance\plengthx by \stemlengthx
\global\advance\plengthy by \stemlengthy
\global\advance\pbackx by \stemlengthx
\global\advance\pbacky by \stemlengthy
\fi  
\global\stemlength=275  \FRONTSTEM=0pt  \BACKSTEM=0pt 
}    
\gdef\drawloop#1[#2#3](#4,#5){  
\input LOOPS  
\drawloop#1[#2#3](#4,#5)}
\Feynmanlength  

\newlength{\extraspace}
\newlength{\extraspaces}
\newcommand{\pr}{\paragraph{}}
\newcommand{\nn}{\nonumber}
\newcommand{\be}{\begin{equation}}
\newcommand{\ee}{\end{equation}}
\newcommand{\bea}{\begin{eqnarray}}
\newcommand{\eea}{\end{eqnarray}}

\begin{document}
\newcommand{\nd}[1]{/\hspace{-0.5em} #1}
\begin{titlepage}

\begin{flushright}
NTUA-57/96 \\
OUTP-96-65P \\

cond-mat/9611072 \\
\end{flushright}

\begin{centering}
\vspace{.05in}
{\Large {\bf Hidden Non-Abelian Gauge Symmetries
in Doped Planar Antiferromagnets \\}}
 
\vspace{.1in}
 K. Farakos 
 
\vspace{.05in}

Physics Department,
National Technical University of Athens, 
Zografou Campus
GR-157 73, Athens, Greece, \\

\vspace{.05in}
and \\
\vspace{.05in}
N.E. Mavromatos$^{*}$ \\
\vspace{.05in}
University of Oxford, Theoretical Physics, 
1 Keble Road OX1 3NP, U.K. \\

\vspace{.2in}
{\bf Abstract} \\
\vspace{.05in}
\end{centering}
{\small We investigate the possibility 
of hidden non-Abelian Local 
Phase symmetries in 
large-$U$ doped planar Hubbard 
antiferromagnets, 
believed to simulate the physics of  
two-dimensional (magnetic) superconductors. 
We present a spin-charge separation ansatz,
appropriate to incorporate holon spin flip, 
which allows for such a hidden local 
gauge symmetry
to emerge in the effective action. The group is  
of the form $SU(2)\otimes U_S(1) \otimes U_{em}(1)$, where 
$SU(2)$ is a local non-Abelian group associated 
with the spin degrees of freedom, $U_{em}(1)$ 
is that of ordinary electromagnetism,
associated with the electric charge of the holes, and 
$U_S(1)$ 
is a `statistical' Abelian gauge group pertaining 
to the fractional statistics of holes on the spatial plane. 
In a certain regime of the parameters
of the model, namely strong  $U_S(1)$ and weak $SU(2)$,
there is the possibility of {\it dynamical formation} of 
a holon condensate.  
This  leads to a 
{\it dynamical} 
breaking of  $SU(2) \rightarrow 
U(1)$.  
The resulting Abelian effective 
theory is
closely related to an earlier model proposed 
as the continuum limit 
of {\it large-spin} planar doped antiferromagnets,
which lead to an unconventional scenario for 
two-dimensional parity-invariant 
superconductivity.}

\vspace{.4in}

\begin{flushleft} 
November 1996 \\
$^{(*)}$~P.P.A.R.C. Advanced Fellow. \\
\end{flushleft} 

\end{titlepage}
\section{Introduction}
\paragraph{}

The discovery of the quasi-planar high-$T_{c}$ Cuprates 
\cite{BM} prompted considerable theoretical interest in
two-dimensional superconductivity of magnetic origin. 
The strong suppression of the isotope effect was one of 
the main reasons for looking for alternatives to phonon mechanisms.
The main feature of the magnetic superconductivity 
was believed to be the fractional statistics of the excitations
on the planar geometry of the materials. 
In two spatial dimensions, particles are
no longer limited to Bose and Fermi statistics
but can acquire an arbitrary interchange phase;
such particles with fractional statistics are known as anyons. Laughlin
\cite{Laughlin}
suggested that a gas of anyons may exhibit superconductivity at low
temperature. This idea was supported by the results of
calculations in the random phase approximation
\cite{FHL}, which demonstrated that a perfect gas of charged anyons with
certain values of the
statistics parameter is indeed a superconductor at zero 
temperature~\footnote{For restrictions on the 
validity of the RPA  approximation 
in the context of 
effective field theories of 
parity-violating 
anyonic superconductivity 
see ref. \cite{kogan}.}. This
`anyonic superconductivity' is an entirely novel phenomenon which has no
analog in three-dimensional systems.
Motivated by the role of anyonic quasi-particles in
the Fractional Quantum Hall Effect,
Laughlin went on to suggest that the charge carriers in the copper oxide
planes
of materials such as La$_{2}$CuO$_{4}$ and YBa$_{2}$CuO$_{6}$
might also have fractional statistics and that superconductivity in these
materials may be well described by the anyonic model.
\paragraph{}
From the experimental point of view, 
however, there seems to be a serious drawback with the anyonic model
as a candidate theory of high-$T_{c}$ superconductivity. A field theoretic
realisation of anyonic matter consists of fermions interacting with an
abelian `statistical'
gauge field whose dynamics is governed by a Chern-Simons term.
As discussed in  \cite{WZ}, this term leads to observable {\it parity violation} 
in an
anyonic superconductor for which there is, as yet, no conclusive
experimental evidence. 
In ref. \cite{dor} 
a proposal was made for a simple
gauge-theory model which 
exhibits two-dimensional superconductivity without
parity violation.
In its most general form, the model consists of
two species of massive fermions coupled with opposite signs
to an abelian gauge field representing effective spin interactions 
among the holon excitations. 
The two species have equal and opposite masses and
hence parity is conserved overall.
This theory may be 
shown~\cite{dor},
to arise as an 
approximate long-wavelength limit of
an idealised model of the dynamics of the charge carriers in 
doped $t-j$ or Hubbard models.
Similar models, but only at a continuum theory level
with no attempt to discuss the connection with 
semi-microscopic condensed-matter systems, 
have been proposed simultaneously in refs. \cite{RK,SW}. 
\paragraph{}
The treatment in ref. \cite{dor} employed 
large-spin approximations for the antiferromagnetic model. 
This resulted in a strong suppression of intrasublattice
hopping~\cite{Sha},
which lead to two species of hole excitations for the 
bi-partite lattices used~\cite{Sha,dor}.
One eventually would like to argue that the same {\it qualitative} 
features occur for the realistic value 
of the spin, $1/2$. 
It is the purpose of this article to attempt to 
formulate the above-mentioned effective theory 
and its physical consequences in a way so as 
to {\it avoid} the large-spin $S$ assumption. 
\pr
To this end, we first review the passage from the 
statistical large-spin models to the continuum 
theories, and then extend the analysis 
to spin $1/2$
models. The local phase symmetries  
that these models possess play a crucial r\^ole in this 
programme, and below we study them in some detail. 
What we shall show here is that the doped
large-$U$ Hubbard models possess
a local $SU(2)\times U_S(1)$ phase symmetry related to spin interactions.
This symmetry will be discovered through a {\it spin-charge 
separation} ansatz, which allows intersublattice 
hopping for holons, and hence spin flip. The spin charge-separation 
may be physically interpreted as implying an effective 
`substructure' of the electrons due to the 
many body interactions in the medium. 
This sort of idea, originating from Anderson's 
RVB theory of spinons and holons~\cite{Anderson},
seems to be pursued recently by 
Laughlin, although from a (formally at least)
different perspective than the one discussed here~\cite{quark}. 
\pr
The effective long wavelength 
model is remarkably similar to a 
three-dimensional gauge model 
of particle physics proposed in ref. \cite{farak}
as a toy example for chiral symmetry breaking 
in QCD. In that work, it has been argued that {\it dynamical 
generation} of a fermion mass gap 
due to the 
$U(1)$ group in $SU(2) \times U(1)$
breaks the $SU(2)$ group down to a $\tau _3-U(1)$ 
group, where $\tau _3$ is the $2\times 2$  Pauli matrix. 
From the particle-theory view point 
this is a Higgs mechanism without an {\it elementary} 
Higgs excitation. 
\pr
The analysis carries over to the present case as well,
if one associates the mass gap to the holon condensate. 
The resulting effective theory of the light degrees of freedom 
is then similar to the 
continuum limit of \cite{dor}
describing unconventional parity-conserving superconductivity.  
\pr
From our point
of view, the above symmetry-breaking pattern summarizes the effects of 
doping on the large-$U$ Hubbard model in a 
dynamical way. In our opinion, 
the appearance of non-Abelian gauge symmetries,
as symmetries of {\it doped} antiferromagnets which 
are broken  
{\it dynamically} by doping, 
and the analogy of the holon-condensate 
formation/superconductivity
with chiral symmetry 
breaking in Yang-Mills theories, 
open up many possibilities for a non-perturbative
(exact) treatment of such theories, including the r\^ole 
of non-perturbative effects in the superconductivity
mechanism~\cite{mavr}. We also believe that our analysis offers 
quantitative support 
to the ideas of refs. \cite{Anderson,quark} about effective
`splitting'
of electrons into spinon and holons in the medium
in a more general context.  
\pr
The structure of the article is as follows: in section 2
we discuss the doped Hubbard (lattice) models from the point of view 
of  
the proposed spin-charge separation ansatz and the associated gauge 
symmetry
structure $SU(2) \times U_S(1)$. In section 3 we discuss the long-wavelength 
effective lattice action in the limit of strong $U_S(1)$, 
the dynamical mass generation for the holes, 
and the connection with 
(Kosterlitz-Thouless) superconductivity. In section 4 
we present an analytical derivation of the dynamical breaking of the $SU(2) \rightarrow U(1)$
on the lattice, in the limit of strong $U_S(1)$. Finally in section 5
we present our conclusions and the possible predictions of the model. 

\section{Hubbard Models and Local Phase
Symmetries}
\subsection{Large-Spin Treatments and their Continuum Limits}
\pr
First let us briefly review the large-spin treatments
of antiferromagnets~\cite{Sha,dor}. 
In the absence of doping impurities, the quasi-planar materials are
antiferromagnetic insulators. 
The potential importance of antiferromagnetic
correlations for high-temperature (cuprate) 
superconductivity was first noted by
Anderson \cite{Anderson} who 
suggested that the correct model for the dynamics
of
electrons in the copper oxide layers was the single-band, large-$U$ Hubbard
model. 
The two-dimensional Hubbard Hamiltonian is written in terms of
operators, $c_{i,\sigma}$ and
$c^{\dagger}_{i,\sigma}$, which annihilate and create
electrons in the $d_{x^{2}-y^{2}}$ orbital at each copper site,
\begin{equation}
H=-t\sum_{\langle ij \rangle , \sigma}c^{\dagger}_{i,\sigma}c_{j,\sigma}
+U\sum_{i}n_{i,\uparrow}n_{i,\downarrow}
\label{Hub}
\end{equation}
where $t$ is the electron-hopping matrix element, $U$ is the strong
Coulomb repulsion and
$n_{i \sigma}=c^{\dagger}_{i \sigma}c_{i \sigma}$ is the occupation
number at each site. In the limit $U\rightarrow\infty$, a
single-occupancy constraint is rigidly
imposed. The undoped case is described by the
Hubbard model with half-filled band and hence the spins are the only
degrees of freedom in this limit. To leading order in large-$U$
perturbation theory, the half-filled Hubbard model is simply equivalent
to the two-dimensional Heisenberg antiferromagnet \cite{Rice};
\begin{equation}
H=J\sum_{\langle ij \rangle} {\bf S}_{i}.{\bf S}_{j}
\label{AF}
\end{equation}
where $J=4t^{2}/U$ and ${\bf S}_{i}$ is the electron spin at site $i$.
Thus, we see that in the infinite $U \rightarrow \infty$ limit 
$J$ corresponds to a weak coupling. 
\paragraph{}
The effective long-wavelength degrees of
freedom of the antiferromagnet can be described by a `relativistic'
quantum field theory in (2+1)-dimensional spacetime. In particular, the
large-$S$ limit of the spin-$S$ Heisenberg  antiferromagnet is
equivalent, at large length-scales, to the quantum nonlinear
$\sigma$-model~\cite{Hal,Aff}.
The relativistic covariance of the effective action
arises from the linear dispersion relation for long-wavelength magnons
and the spin-wave velocity plays the role of the velocity of light in
this formulation. 
\paragraph{}
Doping
introduces mobile charges which hop from site to site against the
antiferromagnetic background of the spins. The coupled
dynamics of holes and spins in the doped
system is highly non-trivial. The hopping of holes tends to disorder the
spins reducing the antiferromagnetic correlation length and the spins
also mediate interactions between the holes. Roughly speaking there is
competition
between the influence of the spins which favour a N\'{e}el-ordered ground-state
and that of the holes which tend to form a spin liquid. A general
conjecture is that a superconducting pairing of holes arises out of this
competition~\cite{Wiegmann,Lee}. This has been verified in ref. \cite{dor}, 
in an effective large-spin analysis. In that analysis 
a large-spin $S \rightarrow \infty$ has been employed, which leads 
to two kinds of holes, due to the suppression of intersublattice 
hopping~\cite{Sha,dor}. To be complete, below we review briefly 
the approach of ref. \cite{dor}, with which we shall make contact
later on. 
\pr
To this end, we 
first note that 
to describe the dynamics of holes in the model of ref. \cite{dor} 
one implements 
a {\it spin-charge separation}, which 
is achieved by representing the electron
operators $c_{i,\sigma}$ using a  `slave-fermion' ansatz
\cite{dorstat,ioff},
\begin{equation}
        c_{i,\sigma}^{\dagger}=\psi_{i} z^{\dagger}_{i,\sigma}
\label{ansatz}
\end{equation}
where $\psi$ is a Grassmann field representing the
absence of a spin at a given site (hole) which carries the
electric charge and $z_{i,\sigma}$ is the spin degree of freedom,
which can be identified \cite{dorstat} with the magnon field of
the $CP^{1}$ $\sigma$-model. The ansatz (\ref{ansatz}) carries
information about a local gauge invariance of
the model as one can perform simultaneous local phase rotations on
$\psi_{i}$ and $z_{i,\sigma}$ fields  with opposite couplings without changing
$c_{i,\sigma}$. It is this symmetry that is responsible for the
gauge nature of the interactions between holes. The physical
reason for such a symmetry is the restriction
of having at most one electron per lattice site.
This redundancy of degrees of freedom is the
characteristic feature of gauge models. 
\pr
The full
partition function of the model is
given as a path-integral over the Grassmann
fields $\psi_{i}$ and $\psi^{\dagger}_{i}$ 
as well as the $CP^{1}$
variables $z$, $\overline{z}$ and $a_{\mu}$.
The corresponding 
long-wavelength
limit 
is derived by linearizing the energy 
spectrum about the chemical potential~\cite{Sha,dor}, and is given by:
\begin{equation}
S_{eff}=\int_{0}^{\beta}d\tau
\int d^{2}x  \frac{1}{\gamma} |(\partial_\mu +ia_\mu) z|^{2}
 + \overline{\Psi}_{a} (i\nd{\partial} - \tau_{3}\nd{a}
+ \frac{e}{c}\nd{A} ) \Psi_{a}
\label{final}
\end{equation}
where $\Psi $ are {\it four-component} Dirac fermions, representing 
holes, $\gamma$ is a constant inversely proportional 
to the $J$ Heisenberg interaction~\cite{dor,dorstat},  
$e$ is the electric charge, $c$ is the velocity of light 
in units of the hole fermi veloccity $v_F=1$, 
and $A$ is an external electromagnetic field. 
The Dirac nature of the holes is a result of the flux-phase assumption
for each sublattice~\cite{dor}.
The reducible four-component
representation of the Dirac algebra in space-time 
is a result of doubling, and  
follows directly from the
local sublattice structure defined by the antiferromagnetic order.  
The opposite statistical charges of 
the holes in different sublattices leads to a $\tau_{3}$
coupling for the gauge field, where $\tau_3 =\left(\begin{array}{cc}
 1 \qquad 0 \\0 \qquad -1 \end{array}\right)$, is the 
$2\times 2$ 
Pauli $\sigma _3$ matrix representation for 
the generator of the
$\tau_3-U(1)$ group~\cite{dor}~\footnote{Here and in ref. \cite{dor}
we follow the terminology $\tau_3-U(1)$ for the abelian group in
spin or sublattice space 
generated by the $2\times 2$ $\sigma_3$ Pauli matrix, so as 
to distinguish it from space-time groups.}.  
\paragraph{}
Integrating 
out the electrically-neutral magnon fields, 
and keeping only the leading terms in a
derivative expansion~\cite{Polyakov,IanA}, 
one obtains 
the low-energy 
effective action of the electrically charged degrees of freedom:
\begin{equation}
{\cal
L}=-\frac{1}{4g^{2}}f_{\mu\nu}f^{\mu\nu}+\overline{\Psi}_{a}(i\nd{\partial}-
S\tau_{3}\nd{a} -\frac{e}{c}\nd{A})\Psi_{a}.
\label{Laginf3}
\end{equation}
The dimensionful gauge coupling $g^{2}$ is proportional to~\cite{IanA,dorstat}
\be 
g^2 =
(\gamma)^{-1}\sim J\eta
\label{gJ}
\ee
where $\eta$ is the doping concentration in the sample. 
In the context of the $t-j$ model, which was 
considered in ref. \cite{dor}, this coupling may be taken 
strong enough so as to generate dynamically 
a gap in the hole spectrum. 
\pr
The above analysis essentially {\it postulated}
the existence of two holon species, by suppressing 
intra-sublattice hopping. This was the result 
of a large-spin analysis. It is the purpose 
of this article to demonstrate that
{\it qualitatively} similar long-wavelength results
may be obtained for spin-$\frac{1}{2}$ doped antiferromagnets.
An important tool in such an analysis is the study of 
local phase symmetries
of the model, which we now turn to. We shall start with a 
review of (non-Abelian) gauge symmetries that characterize
the half-filled (undoped) models, and then proceed
to a study of the doped case upon constructing an 
appropriate {\it spin-charge separation} ansatz, extending 
(\ref{ansatz}) appropriately so as to allow 
intersublattice hopping of holons. 
As we shall show, under the 
proposed ansatz, the effective Hamiltonian 
of {\it holon} and {\it spinon} degrees of freedom  
is characterised by {\it hidden} 
non-Abelian local phase symmetries.  
However, the holon condensate {\it breaks} the 
non-Abelian symmetry 
{\it dynamically}
down to the abelian subgroup discussed in ref. \cite{dor}, and 
hence one recovers the above-discussed Abelian model as 
an {\it effective} theory of the light degrees of freedom. 
Nevertheless, there are {\it remnants} of the non-Abelian symmetry structure,
which manifests itself in the (mass) spectrum of meson-like
excitations as we shall discuss in section 3. 
The presence of such excitations 
constitutes physically testable {\it predictions} of the 
spin-charge separation ansatz proposed in this work.

\subsection{Half-filled Spin-$\frac{1}{2}$ Antiferromagnets:
SU(2) Gauge Symmetry Structure} 
\pr
The large-$U$ (Mott) limit of the half-filled Hubbard model
with $j=4t^2/U$ is the Heisenberg model (\ref{AF}).
In ref. \cite{AM} it has been observed that 
{\it in this limit} there is a local $SU(2)$ symmetry associated
with the spin-$1/2$ algebra of the electrons. 
Indeed, using electron operators $c_i^\alpha$ at a site $i$,
corresponding to spin components up or down, $\alpha =1,2$, 
one may represent the Hamiltonian (\ref{AF}) 
as 
\begin{equation}
H=J\sum_{\langle ij \rangle} 
(c^{\dagger,\alpha}
\sigma _{\alpha}^{\beta} c_\beta)_{i}.
(c^{\dagger,\alpha}
\sigma _{\alpha}^{\beta} c_\beta)_{j}
\label{AF2}
\end{equation}
with {\it the constraint of one electron per site}: 
\be 
c_i^{\dagger,\alpha}c_{i,\alpha}=1
\label{constraint}
\ee
$H$ is invariant under the usual global 
$SU(2)$ transformations of the spin-$\frac{1}{2}$
algebra, $c_\alpha \rightarrow c_\beta g^\beta_\alpha $,
with $g_\alpha^\beta$ an $SU(2)$ matrix. 
In ref. \cite{AM} 
a second $SU(2)$ was constructed out of the 
doublet of creation operators 
$(c_2^\dagger, -c_1^\dagger) $. 
Combining these two doublets in a $2 \times 2$ 
matrix
\be
\chi _{\alpha\beta} = \left(
\begin{array}{cc}
c_1 \qquad c_2 \\
c_2^\dagger \qquad -c_1^\dagger \end{array}
\right)
\label{doublet}
\ee
one observes that in addition to the 
{\it global} $SU(2)$ transformations 
$\chi _{\alpha\beta} \rightarrow \chi _{\alpha\gamma}g^\gamma_\beta$,
one can~\cite{AM} define a {\it local }
SU(2) by left multiplication 
\be
     \chi _{\alpha\beta} \rightarrow h_\alpha^\gamma \chi_{\gamma\beta}
\label{localsu2}
\ee
This local symmetry commutes with the global SU(2) mentioned above. 
Writing the global SU(2) spin operators {\bf S} appearing in (\ref{AF}) 
in terms of $\chi $ as
{\bf S}$\propto tr\chi^\dagger \chi {\bf \sigma}^T$, with $T$ denoting 
matrix transposition, one 
can easily see that the Heisenberg interaction (\ref{AF2}) 
is {\it invariant} under this {\it local} $SU(2)$, 
which is thus the symmetry of the large-$U$ Mott limit 
of the {\it half-filled} Hubbard model. 
It should be stressed of course that 
this is {\it not} an exact symmetry of the Hubbard model. 
As shown in ref. \cite{AM}, 
the very constraint (\ref{constraint}) of one electron per site,
which in terms of $\chi$ variables is expressed as 
\be
    Tr\chi ^\dagger \sigma ^3 \chi =0
\label{constr2}
\ee 
results in a time-dependent local gauge symmetry, when combined
with the kinetic term in the Lagrangian 
\be
    L=\frac{1}{2} \sum _{i} tr\chi _i^\dagger (i\frac{d}{dt} 
+ A_{0,i})\chi _i - H 
\label{lagrange}
\ee
where $A_{0,i}$ acts as a Lagrange multiplier implementing
the constraint, and it may be thought of as the third (temporal)
component of the local $SU(2)$ gauge field~\cite{AM}.
Such gauge symmetries 
appear as a general property of the 
Gatzwyler projection of one electron per site,  
due to the fact that such projections are associated with a sort 
of particle-number conservation. 
This local gauge symmetry connects various mean field 
limits of the half-filled Hubbard model~\cite{AM2}. 
\pr
To understand the formal meaning of the above symmetry, 
we return to the $CP^1$ $\sigma$-model, which is supposed 
to describe the low-energy physics of the half-filled
Hubbard model in a bosonized 
framework for the spin excitations. 
We recall that upon resolving the 
constraint ${\overline z} z = 1$, with $z=(z_1,z_2)$ 
a complex SU(2) doublet with boson statistics, 
the $z$ field can be written as a $2\times 2$ unitary matrix:
\be
z=\left(\begin{array}{cc} z_1 \qquad -{\overline z}_2 \\
z_2 \qquad {\overline z}_1 \end{array} \right) =exp(i\xi_a\sigma_a) 
\label{cp1z}
\ee
were $\sigma _a$,$a=1,2,3$ 
are the Pauli generators of $SU(2)$, 
and the {\it real} fields $\xi _i$ are dynamical. 
The gauged $\sigma$-model action in this representation 
reads 
\be
      S_z=\int d^3x \gamma _0^{-1} tr|(\partial _\mu -ig B_\mu)z|^2 
\label{smodel}
\ee
where $\gamma _0$ is a bare coupling constant.
In this representation one is free to {\it gauge} the full $SU(2)$ 
local gauge group in the $\sigma$-model action,
in which case $B_\mu = B_\mu^a \sigma _a,~a=1,2,3$, 
or its Abelian $U(1)$ subgroup $B_\mu = B_3\sigma _3 $.
The action (\ref{smodel})
reads
\be
S=\int d^3x \gamma _0^{-1} [~\sum _{a} (\partial _\mu \xi _a)^2 
+ g^2 B_\mu ^2 + \sum _{a} B_\mu^a (-2g\partial _\mu \xi_a)~]
\label{resol}
\ee
Technically the above representation separates the Goldstone 
modes from the rest of the fields relevant at low momenta~\cite{huerta}.
The resolution of the constraint implicit in (\ref{cp1z}) 
results in a standard mass term for the gauge field $B$,
instead of the quartic coupling ${\overline z}B^2z$.  
\pr
The possibility of gauging the full $SU(2)$ group in the 
$\sigma$ model is equivalent to the local $SU(2)$ 
symmetry of the Heisenberg action (\ref{AF}) 
found in ref. \cite{AM}, given that 
at half-filling only spin excitations (magnons) exist~\cite{Anderson}.
Of course, the equivalence is understood in terms 
of bosonization, which in $2+1$ dimensions,
unlike $1+1$ dimensions, cannot be expressed  
in a closed form, but only as an effective derivative expansion. 

\subsection{Doped spin-$\frac{1}{2}$ Antiferromagnet
and Non-Abelian Gauge Symmetry Structure}
\pr
Doping is expected to break the $SU(2)$ symmetry
between creation and anihilation pairs of electron operators.  
Naively speaking, a spatial hopping term of the form 
$c^\dagger_{\alpha,i} c_{\alpha,j}$ 
does not seem to be invariant under the 
local SU(2) (\ref{localsu2}). 
Away from half-filling one would expect that only 
a local $U(1)$ can survive, which in view of our 
{\it spin-charge 
separation} 
ansatz (\ref{ansatz})  
seems to be the Abelian subgroup of
SU(2) associated with 
$\tau_3$. This local subgroup is the one gauged in the 
$CP^1$ $\sigma$ model, 
and also the one associated 
with the (Berry phase) term describing 
static holes in the model of ref. \cite{dor}. 
In this article we shall present a {\it dynamical}
scenario by which the above symmetry breaking is achieved. 
The scenario will be remarkably similar to a 
three-dimensional 
particle-physics toy model for chiral symmetry breaking 
in QCD~\cite{farak}.
\pr
The key point is to try to uncover the local $SU(2)$ symmetry 
in the doped case by generalizing the
spin-charge separation ansatz (\ref{ansatz}). 
We seek a representation of the spin-charge 
separation that will allow spin flip, but would still
treat the holons as `blind' to the electronic sublattice 
structure. 
To this end, we {\it propose} to represent the 
holon degrees of freedom 
as two-component spinors in a two-dimensional `colour' space, 
representing Dirac spin components, 
$(\psi_1, \psi_2)$, whilst the spin excitations
are represented by the $CP^1$ doublets $(z_1, z_2)$
living in the {\it same} `colour' space. However 
we amend our construction with 
a {\it spin-flip operation}, which, for the 
$z$-magnon degrees of freedom is 
represented 
by the conjugate doublet $(-{\overline z}_2, {\overline z}_1 )$. 
Thus the electron anihilation operators can be expressed as 
\be
    c_1 =  (\psi _1 \qquad \psi _2 ) \left(\begin{array}{c} z_1 \\ z_2
\end{array}
\right) \qquad     c_2 =  (\psi _1 \qquad \psi _2 ) \left(\begin{array}{c} 
-{\overline z}_2 \\ {\overline z}_1 \end{array}\right)
\label{c1c2}
\ee
while 
the correspondiing creation operators can be obtained 
by -$c_2^\dagger $, $c_1^\dagger $, 
with $\dagger$ denoting 
hermitean conjugation.
We believe that this ansatz 
captures the qualitative features behind the 
RVB idea of Anderson~\cite{Anderson} on spinon and holons.  
Essentially (\ref{c1c2}) implies that to anihilate 
an electron with, say, spin up one has to remove 
all the components of the spin. The spin-charge separation 
ansatz implies that to some extent the 
holes should be `blind' to the spin of the electron
(sublattice structure of the antiferromagnet).
This is correctly captured in (\ref{c1c2}), since  
the hole-`spinors' in colour space are the {\it same} for 
both electron components, whilst the (magnon) $z$-doublets 
differ by a spin-flip operation defined above. 
\pr
Technically, it is convenient to  
combine the creation and anihilation operators, following 
the treatment of the half-filled case (\ref{doublet}).
To this end, we {\it propose} that for the large-U limit of the {\it doped} 
Hubbard model the following {\it spin-charge separation} 
ansatz occurs at {\it each site} $i$: 
\be
\chi _{\alpha\beta,i} = 
\psi _{\alpha\gamma,i}z_{\gamma\beta,i} \equiv \left(
\begin{array}{cc}
c_1 \qquad c_2 \\
c_2^\dagger \qquad -c_1^\dagger \end{array}
\right)_i = 
\left(\begin{array}{cc}
\psi_1 \qquad \psi_2 \\
-\psi_2^\dagger \qquad \psi_1^\dagger \end{array}
\right)_i~\left(\begin{array}{cc} z_1 \qquad -{\overline z}_2 \\
z_2 \qquad {\overline z}_1 \end{array} \right)_i 
\label{ansatz2}
\ee
where the fields $z_{\alpha,i}$ obey canonical {\it bosonic} 
commutation relations, and are associated with the
{\it spin} degrees of freedom,  
whilst the fields 
$\psi _{a,i},~a=1,2$ have {\it fermionic}
statistics, and are assumed to {\it create} 
{\it holes} at the site $i$ with spin index $\alpha$. 
They obey the anticommutation relations:
\begin{eqnarray}
\{\psi_{i,\alpha},\psi^{\dagger}_{j,\beta}\}=\delta_{ij} 
\delta_{\alpha\beta}& &
\{\psi_{i,\alpha},\psi_{j,\beta}\}=
\{\psi^{\dagger}_{i,\alpha},\psi^{\dagger}_{j,\beta}\}=0
\label{anticom}
\end{eqnarray}
The ansatz (\ref{ansatz2}) 
has spin-electric-charge separation, since only the 
fields $\psi$ carry {\it electric} charge.
From now on,  
we shall refer to $\psi _\alpha $ as the `holons', 
and to ${\overline z}_\alpha$ as (bosonized) `spinons'. 
The ansatz (\ref{ansatz2}) is an obvious 
generalization of 
(\ref{ansatz}) if one allows intersublattice hopping. 
\pr
It worths noticing that the anticommutation relations 
for the electron fields $c_\alpha$,$c_\beta^\dagger$,
do not quite follow from the ansatz (\ref{ansatz2}). 
Indeed, assuming the canonical 
(anti) commutation relations for $z$ ($\psi$) fields, 
one obtains 
from (\ref{ansatz2}) 
\bea
&~&\{ c_{1,i}, c_{2,j} \} \sim 2 \psi_{1,i}\psi _{2,i}\delta_{ij} \nn \\
&~&\{ c_{1,i}^\dagger, c_{2,j}^\dagger \} \sim 2 \psi_{2,i}^\dagger
\psi _{1,i}^\dagger \delta_{ij} \nn \\ 
&~&\{ c_{1,i}, c_{2,j}^\dagger \} \sim \{ c_{2,i}, c_{1,j}^\dagger \}
\sim 0 \nn \\
&~&\{ c_{\alpha,i}, c_{\alpha,j}^\dagger \} \sim 
\delta _{ij}\sum_{\beta =1,2} [z_{i,\beta} {\overline z}_{i,\beta} + 
\psi_{\beta,i}\psi _{\beta,i}^\dagger],~~\alpha=1,2
\qquad {\rm no~sum~over~i,j}
\label{ccom}
\eea
To ensure {\it canonical} commutation relations for the $c$ operators
therefore we must {\it impose} at each lattice site the 
(slave-fermion) constraints
\bea
&~&\psi_{1,i}\psi_{2,i} = \psi^\dagger _{2,i}\psi^\dagger_{1,i} = 0,\nn \\
&~&\sum_{\beta =1,2} [z_{i,\beta} {\overline z}_{i,\beta} + 
\psi_{\beta,i}\psi _{\beta,i}^\dagger] = 1
\label{imposition}
\eea
Such relations are understood to be satisfied when the 
holon and spinon operators act on {\it physical} states.
Both of these relations are valid in the large-$U$ limit 
of the Hubbard model and encode the non-trivial physics 
of constraints behind the spin-charge separation ansatz 
(\ref{ansatz2}). They express the 
constraint {\it at most one electron or hole per site},
which characterizes the large-$U$ Hubbard models we are considering 
here. From the above analysis, therefore, it becomes clear that 
the ansatz (\ref{ansatz2}) {\it does not} characterize 
a generic 
Hubbard system, but only the appropriate large-$U$ limit, 
where the constraint of one electron per site is valid. 
As we shall discuss in section 4, both of 
the above constraints (\ref{imposition}) are consistent
with the mass spectrum
of the effective long-wavelength theory obtained
from dynamical generation of a fermion condensate. 
\pr
Now let us look at the symmetry structure 
of the spin-separation ansatz
(\ref{ansatz2}), which in view of the previous analysis 
coincides with the symmetry structure of the effective 
large-$U$ Hubbard action. 
First, it appears to have  
a trivial {\it local} 
SU(2) symmetry, if one defines the transformation 
properties of the $z$ fields to be given by left multiplication
with the $SU(2)$ matrices, and those of the $\psi ^\dagger_{\alpha\beta}$
matrices by the left multiplication (\ref{localsu2}).
In this representation, the gauge group 
$SU(2)$ is generated by the $2 \times 2$ Pauli matrices. 
\pr
The ansatz (\ref{ansatz2}) possesses an 
{\it additional}  
local $U_S (1)$ `statistical' phase symmetry, which 
allows fractional statistics of the spin and charge 
excitations. This is an exclusive feature
of the three dimensional geometry. 
This is similar in spirit, although 
implemented in an admittedly less rigorous way, 
to the bosonization technique of the spin-charge 
separation ansatz of ref. \cite{marchetti},
and allows the alternative possibility 
of representing the holes as slave bosons and  
the spin excitations as fermions. 
\pr
In addition, as a consequence of the fact that the fermions 
$\psi$ carry {\it electric charge}, one has an extra 
$U_{em} (1)$ symmetry for the problem. 
\pr
To recapitulate, 
the above analysis, based on the spin-charge 
separation ansatz (\ref{ansatz2}) which allows 
spin flip,  
leads to the following 
local-phase (gauge) group structure for the doped large-$U$ Hubbard model:
\be
  G=SU(2)\times U_S(1) \times U_{em}(1) 
\label{group}
\ee
where the second $U_{em}(1)$ factor refers to electromagnetic 
symmetry due to the electric charge of the holes. 
This symmetry appears as a {\it hidden} symmetry of the
effective holon and spinon degrees 
of freedom obeying the ansatz (\ref{ansatz2}). 
\pr
The presence of the $U_S(1)$
`statistics' changing group factor will be crucial in our analysis.
As we shall 
discuss in the next section, 
in its strong coupling limit it can generate a mass 
gap~\cite{app,kocic,maris}
for the 
fermionic holon fields $\psi$, which for each hole component
breaks parity, thereby producing a statistics changing 
{\it dynamical} Chern-Simons term. 
However, due to the 
even number of fermionic species there is no overall parity 
violation in the model~\cite{dor}. Note that, 
since this 
statistical gauge field couples also to the $z$ fields, their
statistics will be affected as well. 

\subsection{Effective Hamiltonian of the doped Hubbard antiferromagnet} 
\pr
Next,    
we focus our attention in showing 
that the 
various terms in the action 
be expressible
in terms of the $\chi_{\alpha\beta}$ variables,
which would imply that the symmetries 
of the large-$U$ doped Hubbard model 
action, are the symmetries of the ansatz (\ref{ansatz}).
\pr
To this end, we first study the hopping term 
of the dopped hamiltonian, which broke 
explicitly the local $SU(2)$ symmetry 
(\ref{localsu2}) of the electron operators $c_\alpha,c_\beta^\dagger$. 
Let us rewrite this term  
in terms of $\chi_{\alpha\beta}$ variables:
\begin{equation}
H_{hop}=-\sum_{\langle ij \rangle 
}t_{ij} c^{\dagger}_{\alpha,i}c_{\alpha,j}=
-\sum_{\langle ij \rangle }t_{ij}
[\chi^{\dagger}_{i,\alpha\gamma}
\chi_{j,\gamma\alpha}
+ \chi^{\dagger}_{i,\alpha\gamma}(\sigma_3)_{\gamma\beta} 
\chi_{j,\beta\alpha}]
\label{Hop7}
\end{equation} 
where $\sigma _3$ is a $2\times 2$ Pauli matrix, and summation
over the spin indices is implied. 
In terms of the spin and charge 
excitations, appearing in (\ref{ansatz2}),  
then, the hopping term may be written as 
\bea
&~&H_{hop}=-\sum_{\langle ij \rangle }t_{ij}
[{\overline z}_{i,\beta\kappa}{\psi^\dagger}_{i,\kappa\alpha}
\psi_{j,\alpha\gamma}z_{j,\gamma\beta}
+ \nn \\
&~&{\overline z}_{i,\beta\kappa}{\psi^\dagger}_{i,\kappa\alpha}
(\sigma_3)_{\alpha\lambda}\psi_{j,\lambda\gamma}
z_{j,\gamma\beta}]
\label{Hop8}
\eea
and is trivially local-$SU(2)$ symmetric. 
\pr
To complete the analysis  
we should also look at the interaction terms.
The Heseinberg term (\ref{AF2}) can be written 
in the following convenient form~\cite{AM}
\be 
 H=-\frac{1}{8}J\sum_{<ij>} tr[\chi_i \chi^\dagger _j \chi
_j\chi^\dagger_i] 
\label{heisenb}
\ee
which can be linearized in terms of the fermion bilinears
if one introduces in the path integral 
a Hubbard Stratonovich field $\Delta_{ij}$,
in a standard fashion. 
The result of the linearization is: 
\be
H=\sum_{<ij>} tr[(8/J)\Delta^\dagger_{ij}\Delta_{ji} 
+ (\chi^\dagger_i \Delta_{ij} \chi_j + h.c. )]
\label{linearized}
\ee
We then employ the ansatz (\ref{ansatz2}),  
and perform a Hartree-Fock (mean field) approximation 
for the bilinears:
\bea 
&~&<{\overline z}_iz_j> \equiv |A_1|V_{ij}U_{ij} \quad ; \nn \\
&~&<\psi_i^\dagger (-t_{ij}(1 + \sigma_3)+\Delta_{ij})\psi_j>
\equiv |A_2| V_{ij}U_{ij}
\label{gaugelink}
\eea
where, according to the previous discussion, 
we have used the fact that the link variables are 
$SU(2) \times U_S(1)$ group elements, 
due to the specific transformation properties 
of the variables $z$ and $\psi$. 
In the 
above notation $V$ is the $SU(2)$ part and $U$ denotes the Abelian 
$U_S(1)$
group element. 
The amplitudes $|A_i|$, $i=1,2$, of the link variables 
are assumed frozen, as
usual.
By an appropriate normalization of the 
$z$ and $\psi$ fields, this amplitude is common for both link variables,
\be
|A_1|=|A_2|=K
\label{commonampl}
\ee
According to the discussion of ref. \cite{dor}
the amplitude $K$   
is proportional to the Heisenberg exchange interaction $J=4t^2/U$, with 
$t$ the hopping parameter, and also 
to the doping concentration 
in the sample~\cite{dorstat}. We shall come to this issue later on. 
\pr
The result of the Hartree-Fock approximation, then, 
for the combined hopping and interaction  terms in the Hamiltonian 
is: 
\bea
&~&H_{HF}=\sum_{<ij>} tr\left[(8/J)\Delta^\dagger_{ij}\Delta_{ji}
+ (-t_{ij}(1 + \sigma_3)+ \Delta_{ij})(\psi_j         
<z_j{\overline z}_i>\psi_i^\dagger)\right] + \nn \\
&~&\sum_{<ij>}tr\left[{\overline z}_i<\psi^\dagger_i(-t_{ij}(1 + \sigma_3)+ \Delta_{ij})
\psi_j>z_j\right] + h.c.
\label{HF1}
\eea
and using (\ref{gaugelink}),(\ref{commonampl}) one opbtains:
\bea
&~&H_{HF}=\sum_{<ij>} tr\left[(8/J)\Delta^\dagger_{ij}\Delta_{ji}
+ K(-t_{ij}(1 + \sigma_3)+\Delta_{ij})\psi_j V_{ji}U_{ji}\psi_i^\dagger\right] + \nn \\ 
&~&\sum_{<ij>}tr\left[ K{\overline z}_iV_{ij}U_{ij}z_j\right] + h.c. 
\label{HF}
\eea
This is the effective field theory lattice action we propose to 
describe the dynamicsl of the large $U$ Hubbard model. 
It is understood the the constraints (\ref{imposition}) 
should be taken into account, to complement the description. 
It is important to note that the `fermion' fields 
$\psi$ are $2 \times 2$ matrices in the above representation. 
Notice also that the term $t_{ij} (1 + \sigma_3) + \Delta_{ij}$ 
transforms covariantly under a {\it global} $U(1)$ symmetry generated 
by the Pauli matrix $\sigma_3$. This global $U(1)$ symmetry 
acts on the electron operators $\chi_i $ as $\chi_i \rightarrow U\chi_i$,
with $U=e^{i\theta}$, $\theta$ a global phase.  
The $z$-dependent (magnon) terms yield, in the continuum, the 
$CP^1$ $\sigma$-model lagrangian (\ref{smodel})~\cite{dorstat}. 
\pr
In the large-$U$ Hubbard limit, we are considering here, one 
has the folowing order of magnitude estimates:
\be
 J=4t^2/U \qquad t \sim U \eta_{max} \quad ; \quad \eta_{max} << 1
\label{estimates}
\ee
where $\eta_{max}$ is the maximum doping concentration of the sample,
above which superconductivity is destroyed. 
For underdoped cuprates one may consider the case $\eta _{max} << 1$. 
In this limit, one observes from (\ref{HF}) that the Gaussian fluctuations 
of the variable $\Delta_{ij}$ are of order  ${\cal O}(J/|t_{ij}|)$, and 
hence suppressed, as compared to the hopping 
term $t_{ij}$. This means that one may approximate 
$t_{ij}(1 + \sigma_3) 
+ \Delta_{ij} \simeq t_{ij}(1 + \sigma_3) $. Considering the usual case 
with $t_{ij} = t$ for every $i,j$, one may absorb such terms into 
an appropriate rescaling of the fermion fields $\psi$. 
This will be understood 
in what follows. However, we stress once again that in the case of 
finite-$U$ Hubbard models, 
one should consider the effects of the Gaussian variable $\Delta_{ij}$ 
in the lagrangian (\ref{HF}). This will be left for future work.    
\pr
The conventional lattice gauge theory form of the action 
is derived upon  
integrating out the magnon fields, $z$, 
in the path integral. 
As discussed in \cite{IanA,Polyakov},
the 
result of such a path integration of the magnon
fluctuations around the mean field 
yields appropriate Maxwell kinetic terms for the link variable 
$V_{ij}U_{ij}$, which are the dominant 
terms in a low-energy derivative expansion.
The constraint of at most one electron 
per lattice site in (\ref{imposition}) is crucial in such a derivation, 
since its implementation through a Lagrange multiplier field, $\sigma$,
results in a `mass' term for the magnon fields $z$, in the way explained 
in ref. \cite{Polyakov}. The effective Maxwell terms 
in the continuum are of the generic
form: 
\be 
   S_{kin} \propto \int d^3x \frac{1}{\sqrt{\sigma_0}} 
(F_{\mu\nu}^2 + {\cal G}_{\mu\nu}^2) 
\label{Maxwell}
\ee
where $F_{\mu\nu}$, ${\cal G}_{\mu\nu}$ denote the $U_S(1)$ and $SU(2)$  
field strengths respectively, and 
$\sigma_0$ is a vacuum expectation value 
of the Lagrange multiplier field $\sigma$. 
An elementary one-loop renormalization-group 
analysis yields~\cite{Polyakov}: 
\be 
   \sqrt{\sigma_0}=M-4\pi K_R 
\label{transmass}   
\ee
where $M$ is a transmutation mass, and $K_R$ is the 
`renormalized' $K$ coefficient of the $CP^1$ part of the 
action (\ref{HF}). 
From the analysis of ref. \cite{dorstat} we may infer that 
$K_R \propto J\eta$, with $J$ the Heisenberg interaction,
and $\eta$ the doping concentration, which for lightly-doped 
cuprates is $\eta << 1$. 
This implies that the  order of magnitude of 
the coefficient of the Maxwell term (\ref{Maxwell}), resulting from 
the $z$-integration in a derivative expansion, is set by 
the Heisenberg exchange interaction field strength $J$. 
The conventional three-dimensional gauge coupling $g^2$, of 
dimensions $[mass]$,  
is related to $K_R$ by the simple relation (\ref{gJ}):  
\be
1/g^2 \propto K_R^{-1} \propto J^{-1}\eta^{-1}
\label{gKj}
\ee 
Thus, from (\ref{estimates}) one obtains for the {\it dimensionless} 
coupling $g^2a$, with $a$ the lattice spacing of the antiferromagnetic
Hubbard model: 
\be
     \beta_1 \equiv \frac{1}{g^2a} \propto \frac{1}{\eta \eta_{max}^2Ua}  
\label{beta}
\ee
The magnitude of this coupling 
depends on the way the limit $U \rightarrow \infty$ is taken. 
Taking the limit of $U \rightarrow \infty$ such that  
$Ua \eta \eta_{max}^2 >> 1$
one obtains a {\it small} $\beta$, i.e. strong coupling for the 
$U_S(1)$ group. 
The limit of small $\beta$ is 
crucial for the symmetry 
breaking patterns of the non-abelian $SU(2)$ group, as we shall 
discuss in section 3. 
\pr
We now remark that on 
the lattice the kinetic (Maxwell) terms (\ref{Maxwell})
are given by appropriate plaquette terms 
of the form:
\be   
\sum_{p} \left[\beta_{SU(2)}(1-Tr V_p) + \beta_{U_S(1)}(1-Tr U_p)\right]
\label{plaquette} 
\ee
where $p$ denotes sum over 
plaquettes of the lattice, 
and $\beta_{U_S(1)} \equiv \beta_1 $, $\beta_{SU(2)} \equiv \beta_2 = 
4\beta_1$ 
are the inverse square couplings of the $U_S(1)$ and $SU(2)$ groups,  
respectively.
The specific relation between the $SU(2)$ and $U_S(1)$ 
couplings is 
due to the appropriate normalization of the generators
of the groups. 
\pr
At this point  
it is worthy of remarking that for certain Schwinger-Dyson 
treatments of dynamical symmetry breaking~\cite{dor,AM2} a large-$N$ 
treatment is desirable, in which case one assumes that the spin
$SU(2)$ group is replaced by $SU(N)$ with $N$ large enough.  
In that 
case 
the non-Abelian 
coupling is related to the Abelian one through 
\be
\beta_{SU(N)}=2N\beta_{U_S(1)}=2N\beta_1 
\label{betarel}
\ee
This implies that, even in the 
case of strong $U_S(1)$ coupling, 
 $\beta_1 \rightarrow 0$, the large-$N$ (large spin) 
limit may be implemented in such a way so that $\beta_{SU(N)}$ is {\it finite}.
This is the limit of the 
analysis of ref. \cite{farak}. We shall discuss this 
case in section 3, where we shall make contact with the 
results of ref. \cite{dor}, where such a large-$N$ treatment 
had been assumed. 
\pr
Above we did not 
write explicitly the 
chemical potential term $\mu \sum _{i,\alpha} 
c_{i\alpha}^\dagger c_{i\alpha}$ which determines the doping 
concentration in the sample. This term is also expressed in terms 
of the $\chi$ variables, and essentially has the form 
of (\ref{Hop7}) but for $i=j$, which again may be expressed 
in a gauge invariant way upon using the ansatz (\ref{ansatz2}). 
In deriving long-wavelength
continuum limits, one linearizes the energy spectrum about 
the chemical potential~\cite{Sha,dor}. For most of our 
discussion below we shall not write explicitly such terms, as 
they do not affect the symmetry structure of the theory. 

\subsection{Spinor Structure for Holons and Symmetry Breaking 
Patterns}
\pr
Before closing this section we would like to remark that, 
as a result of the $2 \times 2$ 
matrix structure of the fermion fields $\psi$ in (\ref{HF}),  
one may actually change representation 
of the $SU(2)$ group, and, instead of working with $2 \times 2$ 
matrices, one may use a representation 
in which the fermionic matrices $\psi_{\alpha\beta}$ 
are represented as {\it four-component vectors}
(in `colour' (spin) space): 
\be 
  \psi_{\alpha\beta} \rightarrow {\bf \Psi}^\dagger \equiv 
\left(\psi_1~~-\psi_2^\dagger~~\psi_2~~\psi^\dagger _1 
\right)
\label{fourspinors}
\ee 
It is easy to see that 
in this representation the $SU(2)$ group 
is generated by the following matrices: 
\be
\tau_1 = \gamma _3 \equiv  \left(\begin{array}{cc} 0 \qquad {\bf 1} \\
{\bf 1} \qquad 0 \end{array}\right),~  
\tau_2 = \gamma _5 \equiv i\left(\begin{array}{cc} 0 \qquad {\bf 1} \\
{\bf -1} \qquad 0 \end{array}\right),~ 
\tau _3 = \Delta \equiv i\gamma_3\gamma_5 =
\left(\begin{array}{cc} {\bf 1} \qquad 0 \\
0 \qquad {\bf -1} \end{array}\right) 
\label{gammamatr2}
\ee
where the substructures are $2 \times 2$ matrices.
This is the $SU(2)$ representation 
used in ref. \cite{farak} in the context of  
three-dimensional toy models for chiral symmetry 
breaking.  
Remarkably, the same type of symmetry arises in our context between 
creation and annihilation operators 
of holon pairs in the spin-charge separation
ansatz (\ref{ansatz2}). 
\pr
In the analysis of ref. \cite{farak}, to be discussed 
in the context of the present model 
in the next section,
the statistical group $U_S(1)$ 
group is responsible for the dynamical generation of a parity 
conserving mass $<{\overline \Psi} \Psi>$. 
In terms of the dynamical variables
describing creation and annihilation 
of holons, $\psi$, $\psi^\dagger$ 
respectively, the {\it parity conserving} mass depends 
on the holon condensate. 
To see this, it is convenient to 
split the four-component spinors  (\ref{fourspinors})
into two-component ones: 
\be
{\tilde \Psi}_1^\dagger =\left(\psi_1~~-\psi_2^\dagger \right),~~~~
{\tilde \Psi} _2^\dagger=\left(\psi_2~~\psi_1^\dagger \right)
\label{twospinors}
\ee
In this representation  
the two-component spinors ${\tilde \Psi} $ (\ref{twospinors}) 
will act as Dirac spinors, and the $\gamma$-matrix (space-time)
structure will be spanned by the irreducible 
$2 \times 2$ representation. 
The Dirac conjugate field 
${\overline {\tilde \Psi}}$ may be identified directly 
with the hermitean conjugate fields ${\tilde \Psi} ^\dagger$ 
in terms of holon operators. This is due to the fact that in  
a path integral over the holon fields, the 
conjugate fields $\psi ^\dagger$ can be 
considered as {\it independent} degrees of
freedom~\cite{Sha,dor,dorstat}. 
In this representation, the local $SU(2)$ gauge group 
is generated by the familiar $2\times 2$  Pauli matrices $\sigma _a$,
$a=1,2,3$. The parity transformation is defined as
${\tilde \Psi}_1 \rightarrow \sigma_1 {\tilde \Psi}_2$,
${\tilde \Psi}_2 \rightarrow \sigma_1 {\tilde \Psi}_1$,
which in terms of the (microscopic) holon operators $\psi_i, i=1,2,$ 
reads: 
$\psi_1 \rightarrow \psi_2^\dagger, \psi_2 \rightarrow -\psi_1^\dagger$ .  
With these in mind, it is straightforward to observe that the 
parity-conserving mass term 
${\overline {\tilde \Psi}}_1 {\tilde \Psi }_1 
- {\overline {\tilde \Psi}}_2 {\tilde \Psi} _2$ 
can be related to the holon condensation 
\be
   <{\overline {\tilde \Psi}}_1 {\tilde \Psi} _1 - {\overline {\tilde
\Psi}}_2 {\tilde \Psi }_2>
= -2(\psi_1^\dagger \psi_1 - \psi_2^\dagger\psi_2)
\label{holoncond}
\ee
where we took proper account of the 
anticommutation relations (\ref{anticom})
among the grassmann $\psi _{i,\alpha}$,
$\alpha =1,2$. The terms 
$<\psi_\alpha^\dagger \psi_\alpha>$,$\alpha=1,2,$ are holon condensates.
Notice, in the same context, that the 
parity violating mass term $<{\overline {\tilde \Psi}}_1 {\tilde \Psi} _1 
+ {\overline {\tilde \Psi}}_2 {\tilde \Psi} _2>$ 
equals an irrelevant constant,
which may be subtracted. This 
result is consitent with the {\it generic} 
energetics arguments
that disfavor dynamical generation of a parity-violating 
mass in vector like theories with {\it even} flavour number~\cite{Vafa}. 
\pr
The formation 
of holon coondesate due to a statistics changing $U_S(1)$ group
is similar in spirit to the approach of ref. \cite{FHL} 
in the context of the anyonic
superconductivity.
However, as mentioned above, in our case,  due to the 
four-component structure of the fermions, 
there is an even number of fermionic species 
and hence 
no overall parity violation. 
Moreover this mass gap 
is not a singlet under $SU(2)$, as we shall discuss in the next
subsection,
but transforms as a triplet~\cite{farak}, thereby breaking 
$SU(2)$ down to its $\tau_3-U(1)$ subgroup. 
This is the 
$\tau_3-U(1)$ symmetry of the alsatz (\ref{ansatz}),  
leading to the effective action
(\ref{Laginf3}). 
This provides a sort of dynamical breaking of the local 
spin $SU(2)$ group as the result of introducing holes into the system. 
\pr
The breaking of the $SU(2)$ symmetry down to its Abelian 
$\tau_3$ subgroup admits the (physical) interpretation 
of restricting the holon hopping effectively to 
a single sublattice. In a low-energy
effective theory of the massless degrees of freedom 
this reproduces the 
results of ref. \cite{dor,Sha}. 
This scenario
can be readily seen by using the  
four-component spinor representation (\ref{fourspinors}).
Clearly the two off-diagonal 
generators of the $SU(2)$ group (\ref{gammamatr2})
$\gamma _3$ and $\gamma _5$, corresponding to the 
gauge bosons acquiring masses dynamically due to the holon 
condensate, {\it mix} the two sublattices in the notation of 
ref. \cite{Sha,dor}. Indeed, from (\ref{twospinors})
it follows immediately that if a holon of spin, say, $1$ is 
created at a site $i$, these generators would connect it to the 
destruction of a hole with spin $2$ in the neighboring sublattice.
On the other hand, the generator $\Delta $ of the unbroken 
$\tau_3-U(1)$, is block diagonal, thereby not mixing the sublattices. 
The intrasublattice hopping in this approach is then suppressed 
by the mass of the gauge bosons. We are considering here 
the limit of 
{\it infinitely} strong $U_S(1)$~\cite{farak}.  
In such a limit the intra-sublattice 
hopping is completely suppressed, since the 
mass (which is proportional to the infinite condensate)
is infinite~\cite{farak}. This situation, therefore, describes {\it static}
holes. Hole hopping is allowed for strong but finite couplings, 
in which case the holon condensates and masses are finite.
\pr
We shall devote more discussion on
the phase diagram of the
theory, and its comparison 
to that of ref. \cite{dor}, in the next section. 
We would like to 
close this section by noting that, 
in the context of microscopic models of the form (\ref{HF}), 
dynamical formation of holon condensates, 
and hence destruction of antiferromagnetic order,
would occur above a critical doping concentration~\cite{hirsch}. 
To quantify the 
above results on symmetry breaking, 
therefore, one needs proper lattice simulations
of these models. This is left for the future.

\section{Long-wavelength limit of the spin-$\frac{1}{2}$
doped antiferromagnet}

\subsection{Derivation of the Long-Wavelength Hamiltonian}
\pr
We
now proceed in the long wavelength limit of (\ref{HF}),(\ref{plaquette}), 
in the spinor representation for the holon fields, 
discussed in subsection 2.5. To this end,  
we assume -following the analysis of ref. \cite{dor}- 
a non-trivial {\it flux-phase} 
for the gauge field $U_S(1)$. This is crucial in 
yielding 
a {\it Dirac} form for the hole effective 
action~\cite{Burk,AM,dor}. 
The long-wavelength continuum limit 
is then obtained in a similar way as in the abelian 
case of ref. \cite{dor,dorstat}, 
at low energies, 
by linearizing about a specific 
point on the fermi surface~\footnote{In what follows we shall ignore, for simplicity, 
the shape of the fermi surface~\cite{dor}
and therefore deal with conventional relativistic 
lattice models. 
Of course, this will not be the case in a realistic
condensed matter system, where there are known to be large 
fermi surfaces for holes. The relativistic nature 
may be accurate in superconducting models with 
nodes on their fermi surface, when linearization about a node 
is performed. However, for our purposes in this work, 
which are a study of 
the generic symmetry-breaking patterns
of the local group (\ref{group}),
their physical implications 
for superconductivity, 
and 
the connection with the results of ref. 
\cite{dor}, 
such relativistic 
models will be sufficient.}. 
Due to the assumed flux-phase-$\pi$ {\it background}
for the gauge field $U_S(1)$
one gets for the hopping (kinetic) terms 
of the two-spinors (\ref{twospinors})
(ignoring  
interactions for brevity)~\cite{dor}:
\be
  L_{kin} \sim  \sum _{r,\mu} (-1)^{r_{0} + \dots + r_{\mu-1}}
{\overline {\tilde \Psi}}_c(r) {\tilde \Psi} _c(r + {\hat \mu})+ h.c. 
\label{kin}
\ee
where $c=1,2$ is the colour index, not to be confused with 
the space-time (Dirac) index. 
The factors $(-1)^{r_0 + \dots }$ yield a phase $e^{i\pi}=(-1)$ 
per lattice plaquette,
and this result is produced in our case 
by the $U_S(1)$ flux phase background~\cite{dor}.  
As discussed in ref. \cite{Burk}, the form (\ref{kin}) 
corresponds to a Dirac form for the kinetic terms 
of the fermions ${\tilde \Psi }$
upon
making an (inverse) Kawamoto-Smit transformation~\cite{kawamoto}:
\be
        \Psi _c(r) =\gamma_0^{r_0}\dots \gamma_2^{r_2}{\tilde \Psi} _c(r)
\qquad  {\overline \Psi} _c(r) 
={\overline {\tilde  \Psi }}_c(r)(\gamma_2^{\dagger})^{ r_2}\dots 
(\gamma_0^{\dagger})^{r_0}
\label{kawa}
\ee
where $\Psi$ are two-component Dirac spinors, carrying `colour'. 
We stress once again 
that the colour structure is up and above any space-time (Dirac)
structure. Notice that in such a picture 
fermion bilinears of the form  
${\overline \Psi}_{i,c} \Psi _{i,c'}$ ($i$=Lattice
index), for instance 
the condensate (\ref{holoncond}),
are just ${\overline
{\tilde \Psi}}_{i,c} {\tilde \Psi}_{i,c'}$, due to the 
Clifford algebra $\{ \gamma_\mu~,~\gamma _\nu \} =-2\delta _{\mu\nu}$
and (anti)hermiticity properties 
of the $2\times 2$ $\gamma$ matrices
on the Euclidean lattice. This is useful to have 
in mind when we study the spectrum of meson states in 
section 4.      
\pr
In what follows
we shall make use of the above-mentioned (irreducible) $2\times 2$ 
representation 
in both the colour and space-time 
indices {\it on the lattice}.  
According to the above discussion, then,  
upon ignoring for the moment the electromagnetic 
interactions of holes, 
one obtains the following 
effective low-energy lattice action
for the holon fields, originating from (\ref{HF}),(\ref{plaquette}),
(\ref{kin}):
\bea
&~& S=\frac{1}{2}K \sum_{i,\mu}[{\overline \Psi}_i (-\gamma_\mu) 
U_{i,\mu}V_{i,\mu} \Psi_{i+\mu}  + \nonumber \\
&~& {\overline \Psi}_{i+\mu}
(\gamma _\mu)U^\dagger_{i,\mu}V^\dagger_{i,\mu}
\Psi _i ]  \nn \\
&~& + \beta _1 
\sum _{p} (1 - trU_p) + \beta _2 \sum _{p} (1-trV_p) 
\label{effeaction}
\eea
where $\mu =0,1,2$, $U_{i,\mu}=exp(i\theta _{i,\mu})$ 
represents the 
statistical $U_S(1)$ gauge field, 
$V_{i,\mu}=exp(i\sigma ^a B_a)$ is the $SU(2)$ gauge field, 
and the plaquette terms are obtained, at low energies, as a result of  
the $z$-magnon integration~\cite{IanA,Polyakov}~\footnote{We would 
like to 
mention that, technically, in order 
to study dynamical formation of fermion 
condensates on the lattice using Monte-Carlo studies as 
in ref. \cite{farak},
one should add to the action 
(\ref{effeaction})
a bare mass term
$m_0\sum _{i} {\overline \Psi}_i\sigma_3\Psi _i$, and  
take the limit $m_0 \rightarrow 0$ only at the very end 
of the computations. This will be irrelevant for our 
purposes here.}. 
The fermions $\Psi $
are taken   
to be two-component spinors, in both Dirac and 
colour spaces.  
The quantity $K$, is proportional
to the holon hopping matrix element,
which in turn depends~\cite{Anderson,dor} 
on the doping concentration, 
as stated 
eralier (\ref{gKj}).
According to the discussion following (\ref{betarel}),
in a large-spin ($SU(N),~\rightarrow \infty)$) treatment,
as in ref. \cite{dor}, 
the coupling constant
$\beta _2 \rightarrow \infty$, and hence the  
non-Abelian-gauge-group 
sector of the model is {\it weakly} coupled in this limit.
On the other hand, the coupling of the 
statistical $U_S(1)$ is considered as {\it strongly} coupled,
in the limit $U \rightarrow \infty$. 
It is known, from either lattice 
results~\cite{kocic}, or semi-analytic Schwinger-Dyson (SD) type of 
analyses~\cite{app,maris}, that
dynamical mass generation in a $U(1)$ theory in three space-time 
dimensions 
occurs only for strong
coupling,
i.e. for values of the gauge coupling that are larger than a 
given critical value. 
This mass will break the $SU(2)$ gauge group {\it dynamically}. This 
will be discussed in detail in section 4. 
\pr
The above limit has been studied in ref. \cite{farak}, where 
the model (\ref{effeaction}) has been used as a toy model 
for studying chiral symmetry breaking patterns
of QCD. Remarkably, as we have described above, 
this model can also be used to describe the 
physics of the spin-charge separation 
of strongly correlated electrons in a  doped Hubbard 
model in its large-$U$ limit. In this analogy the 
holon fields $\psi _{\alpha\beta}$ behave like 
the `quarks' of $QCD$, which are thus 
viewed as substructures of the 
physical electron $\chi_{\alpha\beta}$. 
It seems to us that this point of view 
is similar in spirit to that pursued in the 
context of anyonic models 
by Laughlin~\cite{quark}. However, we should stress
that from our point of view this `splitting' is viewed as a 
many-body effect for the holon dynamics
in such systems, 
and hence we do not ascribe to it any further 
significance. 
\pr
\subsection{Symmetry Structure in the Continuum} 
\pr
It will be instructive
to study first the symmetry structure 
of the model 
(\ref{effeaction}) in the continuum, 
following the analysis of 
ref. \cite{farak}. This will help the reader 
understand better the interplay between 
the irreducible ($2\times 2$) and the reducible ($4\times 4$)
representations 
of the Dirac and colour (gauged-chiral 
symmetry) groups.
To this end, we first note that 
the continuum limit of the model (\ref{effeaction})
is described by the lagrangian~\cite{farak}:
\be 
{\cal L} = -\frac{1}{4}(F_{\mu\nu})^2 -\frac{1}{4}({\cal G}_{\mu\nu})^2
+ {\overline \Psi} D_\mu\gamma _\mu \Psi -m_0 {\overline \Psi} \Psi 
\label{contmodel}
\ee
with $D_\mu = \partial _\mu -ig_1 a_\mu^S - ig_2 \sigma^a B_{a,\mu} $,
and $F_{\mu\nu}$,${\cal G}_{\mu\nu}$ are the corresponding field
strengths
for the abelian (statistical) gauge field $a^S_\mu$ and 
the spin SU(2) gauge field $B^a_\mu$. 
The
parity conserving bare mass $m_0$ term 
has been added by hand, as mentioned 
above, to facilitate Monte-Carlo studies of 
dynamically generated fermion masses 
as a result of the formation of fermion condensates
$<{\overline \Psi} \Psi >$ by the strong 
$U_S(1)$ coupling. The $m_0=0$ limit should be taken 
at the end.
\pr
To understand better the nature of this $SU(2)$ gauge symmetry, 
it is instructive to look first at the {\it global} $SU(2)$ 
group, whose gauging produces 
the action (\ref{contmodel}). 
To this end we observe that 
the $\gamma _\mu $, $\mu =0,1,2$,  
matrices, which 
span the {\it reducible} $4 \times 4$ representation of the 
Dirac algebra in three dimensions in a fermionic theory with an 
{\it even} number of fermion flavours, assume the form~\cite{app}:
\bea
&~&\gamma ^0=\left(\begin{array}{cc} {\bf \sigma}_3 
\qquad {\bf 0} \\{\bf 0} \qquad -{\bf \sigma}_3 \end{array} \right)
\qquad \gamma ^1 = \left(\begin{array}{cc} i{\bf \sigma}_1 
\qquad {\bf 0} \\{\bf 0} \qquad -i{\bf \sigma}_1 \end{array} \right) \nn \\
&~&\gamma ^2 = \left(\begin{array}{cc} i{\bf \sigma}_2 
\qquad {\bf 0} \\{\bf 0} \qquad -i{\bf \sigma}_2 \end{array} \right) 
\label{reduciblerep}
\eea
where ${\bf \sigma}$ are $2 \times 2$ Pauli matrices
and the (continuum) space-time is taken to have Minkowskian signature.
As well known~\cite{app} there exists two $4 \times 4 $ matrices 
which anticommute with $\gamma _\mu$,$\mu=0,1,2$: 
\be
\gamma _3 =\left(\begin{array}{cc} 0 \qquad {\bf 1} \\
{\bf 1} \qquad 0 \end{array}\right), \qquad 
\gamma _5 =i\left(\begin{array}{cc} 0 \qquad {\bf 1} \\
{\bf -1} \qquad 0 \end{array}\right)
\label{gammamatr}
\ee
where the substructures are $2 \times 2$ matrices.
These are the generators of the `chiral' symmetry for 
the massless-fermion theory 
\bea
&~&   \Psi \rightarrow exp(i\theta \gamma _3) \Psi \nn \\
&~&   \Psi \rightarrow exp(i\omega \gamma _5) \Psi 
\label{chiral}
\eea
Note that these 
transformations do not exist in the fundamental two-component 
representation
of the three-dimensional Dirac algebra, and therefore 
tha above symmetry is valid for theories 
with even fermion flavours only. 
\pr
The set of generators $\{ {\bf 1}, \gamma _3, \gamma _5, 
i\gamma_3\gamma _5 \equiv \Delta \}$ form~\cite{farak} 
a {\it global} $SU(2) \times U(1)$ symmetry. 
The identity matrix ${\bf 1}$ generates the $U(1)$ subgroup, 
while 
the other three form the $SU(2)$ part of the group. 
The currents corresponding to the above transformations 
are~\cite{farak}
\be
   J_\mu^\Gamma = {\overline \Psi} \gamma _\mu \Gamma \Psi 
\qquad \Gamma =\gamma _3,\gamma _5, i\gamma _3\gamma _5 
\label{currents}
\ee
and are {\it conserved} in the {\it absence} of a fermionic {\it mass}
term. 
It can be readily verified that the corresponding charges
$Q_\Gamma \equiv \int d^2x \Psi ^\dagger \Gamma \Psi $ lead
to an $SU(2)$ algebra~\cite{farak}:
\bea
 &~& [Q_3, Q_5]=2iQ_\Delta \qquad [Q_5,Q_\Delta ]=2iQ_3 \nn \\
&~&  [Q_\Delta, Q_3]=2iQ_5 
\label{chargealgebra}
\eea
If a mass term is present then there is an anomaly
\be
       \partial ^\mu J_\mu^\Gamma = 2m {\overline \Psi } \Gamma \Psi 
\label{anomaly}
\ee
while the current corresponding to the generator ${\bf 1}$ 
is {\it always } conserved, even in the presence of a fermion 
mass~\cite{farak}. 
\pr
The bilinears
\bea
&~&{\cal A}_1 \equiv {\overline \Psi}\gamma _3 \Psi,  
\qquad {\cal A}_2 \equiv {\overline \Psi}\gamma _5 \Psi,  
\qquad {\cal A}_3 \equiv {\overline \Psi}\Psi 
\nn \\
&~&B_{1\mu} \equiv {\overline \Psi}\gamma _\mu \gamma _3 \Psi,~
B_{2\mu} \equiv {\overline \Psi}\gamma _\mu \gamma _5 \Psi,~
B_{3\mu} \equiv {\overline \Psi}\gamma _\mu \Delta \Psi,~\mu=0,1,2      
\label{triplets}
\eea
transform as {\it triplets} under $SU(2)$. 
The $SU(2)$ singlets are 
\be 
{\cal A}_4 \equiv {\overline \Psi}\Delta \Psi, \qquad 
B_{4,\mu} \equiv {\overline \Psi}\gamma _\mu \Psi 
\label{singlets}
\ee
i.e. the singlets are the parity violating mass term, 
and the four-component fermion number. 
\pr
In two-component notation for the spinors $\Psi$, 
the above bilinears read~\cite{farak}:
\bea
&~&{\cal A}_1 \equiv -i[{\overline \Psi}_1 \Psi_2 
- {\overline \Psi}_2 \Psi_1] ,  
\qquad {\cal A}_2 \equiv {\overline \Psi}_1 \Psi_2 
+ {\overline \Psi}_2 \Psi_1 ,  
\qquad {\cal A}_3 \equiv {\overline \Psi}_1\Psi _1 
- {\overline \Psi}_2 \Psi_2 ,   
\nn \\
&~&B_{1\mu} \equiv {\overline \Psi}_1\sigma _\mu \Psi _2 +
{\overline \Psi}_2\sigma_\mu \Psi_1 \quad 
B_{2\mu} \equiv i[{\overline \Psi}_1\sigma _\mu \Psi _2 -
{\overline \Psi}_2\sigma_\mu \Psi_1], \quad
B_{3\mu} \equiv {\overline \Psi}_1\sigma _\mu \Psi_1
-{\overline \Psi}_2\sigma _\mu \Psi_2, \nn \\
&~&{\cal A}_4 \equiv {\overline \Psi}_1\Psi_1
+ {\overline \Psi}_2\Psi_2, \qquad 
B_{4,\mu} \equiv {\overline \Psi}_1\sigma _\mu \Psi_1
+ {\overline \Psi}_2\sigma _\mu \Psi_2~, \qquad \mu=0,1,2      
\label{2compbilin}
\eea 
with $\Psi _i$ denoting two-component Dirac spinors.
For later convenience we have passed onto a three-dimensional 
Euclidean 
lattice formalism, in which ${\overline \Psi}$ is identified 
with $\Psi ^\dagger$, c.f. (\ref{twospinors}).
In this convention the bilinears 
(\ref{2compbilin}) are hermitean quantities. 
It is this Euclidean formalism that we shall use 
for our lattice treatment in section 4~\footnote{On the 
continuum, of course, ${\overline \Psi}=\Psi ^\dagger \gamma _0$, 
with 
$\gamma _0$ a $2 \times 2$ Dirac matrix, and the hermiticity 
properties of the bilinears depend on the representation 
of the Clifford algerba chosen~\cite{farak}.}.
\pr
One may gauge the above group $SU(2)$
and arrive at the continuum action (\ref{contmodel}), which as we discussed 
above describes the low-energy continuum field theory 
limit of the large $U$ Hubbard model (\ref{HF}),(\ref{plaquette}).
In this way, as we shall discuss below, one can 
generate the fermion condensate ${\cal A}_3$
dynamically. In this context, energetics 
prohibits the generation of a parity-violating 
gauge invariant $SU(2)$ term~\cite{Vafa}, and so 
a parity-conserving mass term necessarily breaks~\cite{farak}
the $SU(2)$ group down to a $\tau_3-U(1)$ sector~\cite{dor}, generated
by the $\sigma_3$ Pauli matrix in two-component notation. 
\pr
\subsection{Connection with Superconductivity} 
\pr
We now compare the model presented in this article 
and that of ref. \cite{dor},
which is known to exhibit unconventional parity invariant 
superconductivity, upon coupling the system to 
external electromagnetic potentials $A_\mu$.
First we 
note that there is an important physical difference
between the two models, concerning the mechanism for mass generation. 
In our model in this article 
the gauge group 
that generates dynamically the fermion mass term 
is the strongly-coupled statistical $U_S(1)$, while the $\tau_3-U(1)$-remnant
of the 
weakly-coupled SU(2) group is weakly coupled, and as such
incapable of inducing mass generation. 
On the other hand, in ref. \cite{dor} the fermion 
gap that lead to superconductivity was due to the 
$\tau_3-U(1)$ gauge boson. This may lead to important differences 
between the finite-temperature phase-diagrams
of the two models. 
Such studies 
are left for future investigations. 
\pr
Nevertheless, as far as the mechanism of superconductivity 
is concerned, the two models appear to be {\it qualitatively}
similar, and it is in this sense that the large-spin 
treatment of ref. \cite{dor} is justified by the results
of the present work. 
Indeed,
the global $U_{em}(1)$ symmetry, which is a subgroup of 
the local
symmetry of the ansatz (\ref{ansatz2}),
corresponds to the {\it electromagnetic symmetry} in the 
statistical model. This symmetry 
can be gauged by coupling the action (\ref{effeaction}) 
to an {\it external} electromagnetic field on the spatial
plane as in ref. \cite{dor}.

\begin{centering}
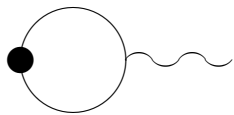
\begin{figure}[htb]
\vspace{2cm}
%
\bigphotons
\begin{picture}(30000,5000)(0,0)
\put(20000,0){\circle{100000}}
\drawline\photon[\E\REG](22000,0)[4]
\put(18000,0){\circle*{1000}}
\end{picture}
%
\vspace{1cm}
\caption{{\it Anomalous one-loop Feynman matrix element,
leading to a Kosterlitz-Thouless-like breaking of the 
electromagnetic $U_{em}(1)$ symmetry, and thus 
superconductivity, once a fermion 
mass gap opens up. The wavy line represents the $SU(2)$ 
gauge boson $B_\mu^3$,
which remains massless, while the blob denotes an insertion 
of the fermion-number
current  $J_\mu={\overline \Psi}\gamma_\mu \Psi$.
Continuous lines represent fermions.}}
\label{fig1}
\end{figure}
\end{centering}

As discussed there, then, 
{\it superconductivity} is obtained upon the 
opening of the gap in the fermion (hole) spectrum
due to the one-loop anomalous effect
corresponding to the following 
Feynman matrix element, depicted in fig. 1: 
\be
    {\cal S}^a = <B^a_\mu|J_\nu|0>,~a=1,2,3~; \qquad J_\mu ={\overline
\Psi}\gamma _\mu \Psi 
\label{matrix}
\ee
with $\Psi$ four-component spinors, which correspond to the
continuum limit of (\ref{fourspinors}). 
It should be stressed  that as a result 
of the colour group structure only the massless $B^3_\mu $ 
gauge boson of the $SU(2)$ group, corresponding to the $\sigma _3$
generator in two-component notation, contributes to the graph. 
The result is~\cite{RK,dor}:
\be
    {\cal S} = <B^3_\mu|J_\nu|0>=({\rm sgn}{M})\epsilon_{\mu\nu\rho}
\frac{p_\rho}{\sqrt{p_0}} 
\label{matrix2}
\ee
where $M$ is the parity-conserving fermion mass 
(or the holon condensate in the context of the 
doped antiferromagnet). 
This observation is consistent with the symmetry-breaking patterns
of the $U_{em}(1)$ group since the $B^3_\mu$ colour component remains 
massless, and therefore plays the r\^ole of the Goldstone boson~\cite{dor}.
As discussed in ref. \cite{RK,dor}, 
this unconventional 
symmetry breaking however does not have a local order parameter,
and thereby resembles, but is not identical to, the
Kosterlitz-Thouless
mode of symmetry breaking~\cite{KT}. The {\it massless} Gauge Boson 
$B_\mu^3$ of the 
unbroken $U(1)$ subgroup of $SU(2)$ is responsible for the 
appearance of a massless pole in the electric current-current 
correlator~\cite{dor}, which is the characteristic feature 
of any superconducting theory. In this sense, in ref. \cite{dor}
the field $B^3_\mu$, or rather its {\it dual} $\phi$ defined by
$\partial _\mu \phi \equiv \epsilon_{\mu\nu\rho}\partial_\nu B^3_\rho$,
was identified with the Goldstone
Boson of the broken $U_{em}(1)$ (electromagnetic) symmetry. In the non-Abelian 
context there are also Goldstone bosons associated with the 
breaking of the $SU(2)$ symmetry~\cite{farak}. These will be discussed
in the next subsection.
\pr

\section{Dynamical Gauge Symmetry Breaking on the Lattice }
\pr
In this section we 
derive the 
symmetry breaking patterns, and discuss, in detail, the 
excitation 
spectrum of the theory
obtained from the effective long-wavelength {\it lattice} action
(\ref{effeaction}). 
We are interested in the effective action
of the holon degrees of freedom, after integrating 
out the fractional-statistics $U_S(1)$ field.
From the above discussion it becomes obvious that this
field plays an auxiliary r\^ole in the spin-separation
ansatz, and as such it should be integrated out
in the effective action of the physical 
degrees of freedom. 
\pr
We shall concentrate on the $\beta _1 =0$ strong
coupling limit for the $U_S(1)$, which from the 
point of view of the doped Hubbard model corresponds to an 
infinite-$U$  limit. 
In this limit the $U_S(1)$ gauge field may be easily integated 
out in the path integral
with the result~\cite{farak}
\be
  \int dV d{\overline \Psi}d \Psi exp(-S_{eff})
\label{action3}
\ee
where
\bea
&~&S_{eff} = \beta _2 \sum _{p} (1 -trV_p)
+ \sum _{i,\mu}{\rm ln}I_0(\sqrt{y_{i\mu}}) \nn \\
&~&y_{i\mu} \equiv 
K^2 {\overline \Psi}_i (- \gamma _\mu) 
V_{i\mu}\Psi _{i+ \mu} {\overline \Psi}_{i + \mu} 
(\gamma _\mu)V_{i\mu}^\dagger \Psi _i
\label{action4}
\eea
and $I_0$ is the zeroth order Bessel function.  
The quantity $y_{i\mu}$ may be written in terms 
of the bilinears
\be
  M^{(i)}_{ab,\alpha\beta} 
\equiv \Psi _{i,b,\beta}{\overline \Psi}_{i,a,\alpha},~~a,b={\rm
colour},~\alpha,\beta={\rm Dirac},~i={\rm lattice~site}
\label{mesons}
\ee
The result is: 
\be
y_{i\mu} = -K^2tr[M^{(i)}(-\gamma _\mu)V_{i\mu}
M^{(i+\mu)}(\gamma_\mu)V_{i\mu}^\dagger] 
\label{y}
\ee
In the analogue
language of particle physics~\cite{farak}
the quantities $M^{(i)}$ 
would represent physical {\it meson} states. 
In the context of our spin-charge
separation ansatz the mesons would be 
composite states of holons. We have already seen that 
the
physical electrons are composites of magnon-holons.
In the theory (\ref{effeaction}) the magnon degrees
of freedom have been integrated out.
In this context, the low-energy (long-wavelength)
effective action is written as a path-integral 
in terms of gauge field and meson states~\cite{farak}
\be
Z=\int [dV dM]exp(-S_{eff}+ \sum_{i}tr{\rm ln}M^{(i)})
\label{effmeson}
\ee
where the meson-dependent term in (\ref{effmeson}) 
comes from the Jacobian in passing from 
fermion integrals to meson ones~\cite{kawamoto}.
\pr
In ref. \cite{farak} a method was presented for 
identifying the symmetry-breaking patterns of the 
gauge theory (\ref{effeaction}), by studying  
the 
dynamically-generated mass spectrum. The method consists
of first expanding $\sum _{i,\mu}{\rm ln}I_0(\sqrt{y_{i,\mu}})$
in powers of $y_{i\mu}$, and concentrating on the lowest 
orders, which will yield the gauge boson 
masses, whilst higher orders describe interactions. 
Keeping only the linear term in the expansion
yields~\cite{farak}
\be
   {\rm ln}I_0(\sqrt{y_{i\mu}}) \simeq 
\frac{1}{4}y_{i\mu} =
-\frac{1}{4}K^2 tr[M^{(i)}(-\gamma _\mu)V_{i\mu}
M^{(i+\mu)}(\gamma _\mu)V_{i\mu}^\dagger]
\label{fkresult}
\ee
It is evident that symmetry-breaking patterns 
for $SU(2)$ will emerge out of a non zero VEV 
for the meson matrices $M^{(i)}$.
\pr
Lattice simulations of the model 
(\ref{effeaction}), with only 
a global $SU(2)$ symmetry, in the strong $U_S(1)$ coupling
limit $\beta _1=0$, and 
in the quenched approximation for fermions, 
have shown~\cite{koutsoumbas} that 
the states generated by the bilinears 
${\cal A}_1$ and ${\cal A}_2$ (c.f. (\ref{triplets}))
are {\it massless}, and therefore correspond to 
Goldstone Bosons, while the state generated 
by the bilinear ${\cal A}_3$ is massive. 
In the context of our statistical model (c.f. (\ref{twospinors}))
these meson states may be expressed in terms of the 
holon operators as:
\bea
&~&{\cal A}_{1,i} = -i({\overline \Psi}_1 \Psi _2 - 
{\overline \Psi}_2 \Psi _1)_i
= -2i(\psi ^\dagger _1 \psi_2 - \psi ^\dagger _2 \psi_1)_i \nn \\
&~&{\cal A}_{2,i} = ({\overline \Psi}_1 \Psi _2 + 
{\overline \Psi}_2 \Psi _1 )_i =-2
(\psi ^\dagger _1 \psi_2 + \psi ^\dagger _2 \psi_1)_i 
\label{a12}
\eea
and the bilinear 
${\cal A}_3$ 
is given by (\ref{holoncond})
\be
{\cal A}_{3,i} =({\overline \Psi}_1 \Psi _1 - {\overline \Psi}_2 \Psi _2)_i
= -2(\psi _1^\dagger \psi_1 - \psi_2^\dagger \psi_2)_i
\label{a3}
\ee
The fact that 
members of the triplet SU(2) representation acquire
different masses is already evidence for 
symmetry breaking. We shall confirm this explicitly 
later on. For the moment we note that 
lattice analyses~\cite{koutsoumbas,kawamoto2} show that  
in the strong coupling limit $\beta _1 =0$ the 
condensate $u \equiv <{\cal A}_3>$ and the mass of ${\cal A}_3$ 
are {\it infinite}. Of course the masses and the condensate 
are finite for finite $\beta_1$, which is the case of finite-$U$ 
Hubbard models (c.f. (\ref{beta})). In addition, in this approximation 
this is the only meson state that develops a {\it non-zero}
vacuum expectation value (VEV). This therefore constitutes
a prediction for the infinite $U$ Hubbard model and 
the spin-separation ansatz (\ref{ansatz2}). 
The fact that the VEV of the Goldstone Boson states 
${\cal A}_{1,2}$ {\it vanish} implies the absence 
of a `spin-flip' (on the average) 
at a 
site: $<\psi_{1,i}^\dagger \psi_{2,i}>=<\psi_{2,i}^\dagger \psi_{1,i}>=0$, 
which is also consistent with 
the slave-fermion {\it constraints} (\ref{imposition}).  
This is also comforting from the point of view of the 
equivalence of the above $U \rightarrow \infty$ Hubbard 
model with that of ref. \cite{dor}, 
whose symmetry breaking dynamical 
patterns are characterized by 
the absence
of a local order parameter~\footnote{The absence 
of VEVs for the Goldstone Bosons ${\cal A}_{1,2}$
eliminates 
a potentialy dangerous source of a possible appearance 
of a local order parameter in the model. 
Notice that the dynamical 
breaking of the electromagnetic $U_{em}$ symmetry 
as a result of the holon condensate
occurs without  
a local order parameter~\cite{dor}.}. 
\pr
One has 
the following expansion for the 
meson states in terms of the $SU(2)$
bilinears  (\ref{2compbilin})~\cite{farak}:
\bea
 &~& M^{(i)} = {\cal A}_3(i)\sigma_3 
+ {\cal A}_1(i)\sigma_1 + {\cal A}_2 (i)\sigma_2 + 
{\cal A}_4 (i){\bf 1} + \nn \\
&~& i[B_{4,\mu}\gamma ^\mu + B_{1,\mu}(i)\gamma^\mu\sigma_1
+ B_{2,\mu}\gamma^\mu\sigma_2 + B_{3,\mu}\gamma ^\mu \sigma_3 ] 
\label{mesontripl}
\eea
with $\mu=0,1,2$,
$\gamma_\mu$ are (antihermitean) Dirac (space-time) $2\times 2$ matrices,
and $\sigma_i$, $i=1,2,3$ are the (hermitean) $2\times 2$ 
SU(2)-`colour' Pauli matrices. Note that 
the VEV of the matrix $<M^{(i)}>=u\sigma_3$
is proportional to the chiral condensate. 
Upon substituting (\ref{mesontripl})
in (\ref{fkresult}), taking into account that the 
SU(2) link variables may be expressed as: 
\be 
V_{i\mu} = cos(|{\bf B}_{i\mu}|) + i{\bf \sigma}.{\bf B}_{i\mu}
sin(|{\bf B}_{i\mu}|)/|{\bf B}_{i\mu}|
\label{linkexpres}
\ee
and performing a naive perturbative expansion
over the fields ${\bf B}$ one finds:
\be
{\rm ln}I_0(\sqrt{y_{i\mu}}) \propto 
K^2 u^2
[(B^1_{i\mu})^2 + (B^2_{i\mu})^2]~+~{\rm interaction~terms}  
\label{massiveboson}
\ee
From this it follows that two of the $SU(2)$ gauge bosons, 
namely the $B^1$,$B^2$ become {\it massive}, with masses
proportional to the chiral condensate $u$: 
\be
{\rm B^{1,2}~boson~masses} \propto K^2u^2
\label{massbooson}
\ee
whilst the gauge boson $B^3$ remains {\it massless}.
\pr
This mass term breaks $SU(2)$ to a $U(1)$ subgroup, 
and in view of the above analysis one recovers the
effective action for the massless modes occuring in the  
large-spin treatment of ref. \cite{dor}, and reviewed 
in section 2.  
It is understood 
that a full analysis for finite values
of $\beta _1$ is necessary, before definite conclusions
are reached in connection with the exact properties and 
physical implications of the ansatz (\ref{ansatz2}) 
for finite $U$ doped Hubbard, or $t-j$, models. 
We hope to come back to these issues in the future. 
\pr
We would like now to draw the reader's attention to the 
similarity of the above mechanism for symmetry breaking 
with the situation 
in the adjoint gauge-Higgs model~\cite{adjoint}. 
There, the $SU(2)$ symmetry is also broken down to a $U(1)$
whenever the constant multiplying the Higgs-gauge interaction 
is larger than a critical value. 
In our case  
the r\^ole of this constant is played by $K^2$, 
as can be seen by the formal analogy between 
the adjoint-Higgs-gauge interaction terms and (\ref{fkresult}). 
Of course, in our
approach symmetry
breaking was achieved due to    
the infinitely strong $U_S(1)$ coupling. 
In view of the above analogy with the 
adjoint-Higgs model~\cite{adjoint}, however, 
one may speculate that interesting 
phase diagrams for the symmetry breaking of $SU(2)$ 
could also emerge due to the $K^2$ coupling, 
in a way 
independent of the $U_S(1)$ coupling. 
In this respect, we would like 
to stress once again 
that in the context of our statistical models~\cite{dorstat}
the amplitude $K$ is proportional 
to the doping concentration in the sample,
$K \propto J\eta$. 
Since the adjoint-Higgs-like symmetry breaking 
requires 
strong enough coupling, then the above analysis,
if true in this context, 
may be seen to  
suggest 
a natural and simple explanation 
-in the context of a gauge theory - of 
the fact that in 
planar antiferromagnetic models
of finite-$U$-Hubbard or $t-j$ type,   
antiferromagnetic order is destroyed, in favour of superconductivity, 
{\it above} a critical doping concentration. As mentioned
at the end of section 2, this point of view seems to be supported
by preliminary results of 
lattice simulations~\cite{hirsch}. More detailed 
investigations along this line of thought 
are left for future work. 

\section{Conclusions  and Outlook}
\pr
In this article we have discussed lattice models for 
{\it planar} 
spin-$\frac{1}{2}$ Heisenberg Antiferromagnets
away from half filling (doped). We have worked in the
infinite $U \rightarrow \infty$ limit of the Hubbard model,
which is characterized by the Gatzwyler projection, 
namely a constraint of {\it no more than one electron 
per lattice site}. 
Upon implementing a spin-charge separation ansatz (\ref{ansatz2}),
in a way consistent with holon spin flip,  we have argued 
that the {\it doped} model is still characterized by 
a local $SU(2) \times U_S(1) \times U_{em}(1)$ symmetry
upon coupling to external electromagnetic fields. 
Of these, the $U_S(1)$ is an auxiliary `statistical' gauge symmetry,
associated with the fractional statistics of the spin and charge 
excitations in the ansatz (\ref{ansatz2}). This possibility 
arises because of the planar spatial structure of the lattice model. 
\pr
We have argued 
that for strong enough $U_S(1)$ couplings, 
dynamical generation of a holon condensate 
can occur, with the result of breaking the $SU(2)$ 
group to $\tau _3-U(1)$.  This is the same 
local phase symmetry 
as the one characterising 
superconducting effective theories of 
doped antiferromagnets in large-spin $S \rightarrow \infty$ 
treatments~\cite{Sha,dor}, although the mechanisms
for mass generation are different. Nevertheless,
the superconductivity scenaria appear qualitatively similar. 
In this way we have explained
two things in a dynamical way: (i) the {\it breaking}- 
as a result of {\it doping}- of 
the local SU(2) spin symmetry that characterizes 
half-filled large-$U$ Hubbard models, and 
(ii) the qualitative justification of large spin treatments
and in particular the suppression of intrasublattice hopping 
of holes. Indeed, the latter is associated with 
massive SU(2) gauge boson states, which acquire their masses 
through  
holon condensation. There are many features of the models that 
still have to be 
worked out. Finite-$U$ treatments and extension of these 
ideas to $t-j$ models are worth pursuing.  
Given the dependence of the coupling constants of such models 
on the doping concentration in the sample, then 
a
renormalization-group study of the respective phase diagrams
could provide useful quantitative information 
on the order of magnitude of the maximum doping concentration
for superconductivity, and, in general, shed more light on the 
physics of the spin-charge separation in the models. 
We hope to come to 
a more systematic study of such issues in the future.
\pr
Further consistency checks 
of our approach 
may also come from a study of the
renormalization group structure of the 
{\it normal phase} of the model 
in the infrared. By normal phase we mean the phase
where 
there is no dynamical opening of a gap. 
In this respect we mention that 
in three space-time 
dimensions the natural coupling 
constant appearing in the Lagrangian of a $U(1)$ gauge theory with
fermions
is a parameter with dimensions of $\sqrt{{\rm mass}}$. 
In analytic Schwinger-Dyson  treatments one can define a dimensionless 
coupling, which is  
essentially the ratio of the 
coupling constant over a characteristic 
mass scale of the theory, playing the r\^ole of the 
ultraviolet cut-off~\cite{app}. 
In a recent series of papers~\cite{AitchMavr}, it was argued that 
this dimensionless coupling decreases slowly 
with the momentum scale.
Its growth towards the infrared regime, however, 
is
cut-off 
by the appearance of a {\it 
non-trivial infrared fixed point}.  
The latter 
phenomenon is responsible 
for deviations from fermi-liquid 
behaviour~\cite{Shanorm,AitchMavr}, and - if the 
infrared fixed-point value of the coupling is strong enough~\cite{app}-
also for mass generation.   
These features are expected to 
persist in the present model. 
However, in the present case, the full non-Abelian 
$SU(2) \times U_S(1)$ symmetry will be present in the normal phase. 
A full analysis 
along the lines of ref. \cite{AitchMavr} remains to be done. 
\pr
Above we have dealt with 
relativistic low-energy limits, obtained by linearizing about 
specific points  on the fermi surface for the holons. 
As argued 
in ref. \cite{AitchMavr} this may still capture 
certain qualitative features of realistic non-relativistic
holon models. Eventually, one would like 
to be able to extend quantitatively the above results
to non relativistic cases as well.
We mention, however, that our relativistic limits
may be related to 
condensed matter systems with fermi surfaces that have nodes.
Such systems are known to exist in nature, and in particular
they are antiferromagnetic planar systems with a strong 
spin-chain anisotropy as far as Heisenberg interactions 
are concerned~\footnote{We thank A. Tsvelik for useful 
information on the existence of such  
materials.}. Upon doping 
and linearization around holon-fermi-surface nodes,
one might then obtain the effective relativistic models 
discussed in this work and in ref. \cite{dor}. 
\pr
An important issue we would like to raise
as a result of the present work is the fact 
that non-Abelian local gauge symmetries, arising  
in the strong $U$ Hubbard antiferromagnets, 
imply the possibility of existence of non perturbative
effects (monopole-instantons in the form of Hedgehog 
configurations  
etc). 
Their precise r\^ole in the superconductivity mechanism 
associated with these models 
needs to be investigated in detail~\cite{dor,mavr}.
This becomes particularly important in view of the 
claimed association of this scenario for superconductivity
with Kosterlitz-Thouless-like phase transitions~\cite{dor}. 
There are important similarities between the two scenaria, since
both are characterized by the absence of local order parameters
for the Goldstone bosons associated with the symmetry breaking. 
It is known that in Kosterlitz-Thouless transitions
the symmetry breaking occurs when non-perturbative 
degrees of freedom are liberated. A preliminary 
analysis~\cite{dor,mavr} in the effective theory model 
of ref. \cite{dor}, which, as a result 
of the present work, may be  viewed as an effective 
theory of the massless degrees of freedom 
of the non-Abelian case,
has shown that non-perturbative 
effects appear to be bound in 
pairs in the superconducting phase. This issue deserves however
further investigations that require going beyond
perturbation theory. 
\pr
In this latter respect, we mention that the treatment 
of non-perturbative effects requires {\it exact} results.
Of course 
the superconductivity mechanism advocated in ref. \cite{dor}
occurs through an anomaly, which is an exact one loop result.
However, this is not sufficient for an exact quantitative
treatment of the low-energy effective action.
However, it is known that 
exact results 
in effective action treatments
in higher than one spatial dimension
can be derived in 
certain {\it supersymmetric} {\it non-Abelian} gauge  
theories, as a result
of special non-renormalization theorems
and strong-weak coupling duality symmetries~\cite{seiberg}. 
In such theories,  
one invokes a duality symmetry to map 
a strongly-coupled problem to a weakly-coupled
dual model which can be solved exactly. 
\pr
We now remark that
$t-j$ models are known,  
under certain restrictions 
among their parameters - namely $t=j, $ 
to exhibit {\it hidden supersymmetries} in 
space time~\cite{susy}. There are graded algebras
among the three possible states on a lattice site
of the $t-j$ model~\cite{susy}: $|a>=\{ |0>,|1>,|2> \}$,
corresponding to the empty, spin up and spin down states
respectively.  
The model is supersymmetric
up to a shift 
in the chemical potential, in the 
sense that there exist two supercharge
operators $Q^{+}_\sigma $, $\sigma=1,2$ (SU(2) `spin' index),
connecting fermi and bose sectors and leaving the action invariant. 
So far
this supersymmetry structure was not given any
dynamical significance. 
This is because
this supersymmetry refers to electron operators. 
Our ansatz (\ref{ansatz2}), however, 
which implies electron substructure, 
when and if extended 
to this case,
might imply 
hidden supersymmetries
among holon and spinons. These might 
have non-trivial consequences on the dynamics,
following the spirit of ref. \cite{seiberg},
provided one could extend it to this case.  
In such a context, 
the superconductiviity model of ref. \cite{dor}
could be viewed as an 
{\it effective} theory of the light
degrees of freedom, arising 
in the gauge-symmetry-broken phase of a
supersymmetric $SU(2)\times U(1) \times U_{em}(1)$ 
field-theory model of a doped antiferromagnet with  $t=j$.
\pr
At present, we lack any microscopic dynamics
underlying (\ref{ansatz2}) that would allow us to check on 
its generalization to the $t=j$ case and on the 
existence of 
the above-conjectured supersymmetric structure. 
At any rate, we believe that our work of associating 
holon condensation 
with a dynamical breaking of a {\it Yang-Mills} gauge 
theory in {\it doped} antiferromagnetic planar systems
is an interesting observation, which deserves
further serious investigations.
We do hope to come back to a study of 
some of the above-mentioned issues
in due course. 

\setcounter{equation}{0}
\paragraph{}
\paragraph{}
\noindent {\Large {\bf Acknowledgements}}
\paragraph{}
The authors are grateful to Ian Aitchison  for many enlightening 
discussions and a critical reading of the manuscript. 
They would also like to thank I. Kogan, G. Koutsoumbas, 
F. Lizzi and A. Tsvelik
for useful comments,  
and the Theory Division of CERN 
for hospitality during the initial stages of this work.

\end{document}